\documentclass{article}

\usepackage[preprint]{neurips_2022}

\usepackage[utf8]{inputenc} % allow utf-8 input
\usepackage[T1]{fontenc}    % use 8-bit T1 fonts
\usepackage{hyperref}       % hyperlinks
\usepackage{url}            % simple URL typesetting
\usepackage{booktabs}       % professional-quality tables
\usepackage{amsfonts}       % blackboard math symbols
\usepackage{nicefrac}       % compact symbols for 1/2, etc.
\usepackage{microtype}      % microtypography
\usepackage{xcolor}         % colors

\usepackage{amsmath}
\usepackage{amssymb}
\usepackage{mathtools}
\usepackage{amsthm}

\title{Triangulation candidates for Bayesian optimization}

\author{%
  Robert B.~Gramacy\\
  Department of Statistics\\
  Virginia Tech\\
  Blacksburg, VA 24060, USA\\
  \texttt{rbg@vt.edu} \\
  \And
  Annie Sauer\\
  Department of Statistics\\
  Virginia Tech\\
  Blacksburg, VA 24060, USA\\
  \texttt{anniees@vt.edu} \\
  \And 
  Nathan Wycoff\\
  McCourt School of Public Policy\\
  Georgetown University\\ 
  Washington DC, 20057, USA \\
  \texttt{nathan.wycoff@georgetown.edu} \\
}

\begin{document}

\maketitle
\vspace{-0.25cm}  % THIS WAS ADDED FOR PREPRINT (UNBLINDED) VERSION, REMOVE FOR SUBMISSION

\begin{abstract}
\vspace{-0.2cm} % THIS WAS ADDED FOR PREPRINT (UNBLINDED) VERSION, REMOVE FOR SUBMISSION
Bayesian optimization involves ``inner optimization'' over a new-data
acquisition criterion which is non-convex/highly multi-modal, may be
non-differentiable, or may otherwise thwart local numerical optimizers.  In
such cases it is common to replace continuous search with a discrete one over
random candidates.  Here we propose using candidates based on a Delaunay
triangulation of the existing input design.  We detail the construction of
these ``tricands'' and demonstrate empirically how they outperform {\em both}
numerically optimized acquisitions and random candidate-based alternatives,
and are well-suited for hybrid schemes, on benchmark synthetic and real
simulation experiments.
\end{abstract}

\section{Introduction}
\label{sec:intro}

We address the continuous unconstrained optimization problem
\begin{equation}
x^\star = \underset{x \in \mathcal{B}}{\mathrm{argmin}} \; f(x) \label{eq:mp}
\end{equation}
where the bounding box $\mathcal{B}$ is a hyperrectangle, often taken as
$[0,1]^d$ in coded inputs.  The objective $f:\mathcal{B} \rightarrow
\mathbb{R}$ is a {\em blackbox} function, meaning that 
%its operations are generally opaque to us, and 
we can only learn about its behavior through expensive, often
simulation-based, evaluation.  %Often, $f$ is a numerical simulation carried out on a
%computer, where expense is measured in core-hours.  
Such problems are most
challenging when $f$ is highly non-convex, and thus contains multiple local
minima. %Consequently the 
A tacit goal of a solver is to minimize the number of times that $f$ is
evaluated in search of a global solution.

The earliest papers on Bayesian optimization (BO) adapted statistical modeling
and design principles to tackle this optimization problem
\citep{movckus1975bayesian,jones1998efficient}. Applications on
physics-based simulators $f$ are provided by \citet{pourmohamadbayesian};
scenarios in machine learning are reviewed by
\citet{garnett_bayesoptbook_2022}. BO is common in studies of engagement and
user experiences in online platforms \citep[e.g.,][]{letham2019bayesian},
hyperparameter estimation for deep learning
\citep[e.g.][]{turner2021bayesian,feurer2018practical} and materials design
\citep[e.g.][]{zhang2020bayesian}, to name a few.

To illustrate BO and introduce our contributions, consider $f(x) =
\sin(x)$ and $\mathcal{B} = [-1,2\pi+1]$ evaluated at six
equally-spaced inputs $x$.  Next fit a surrogate $\hat{f}_n$, to data $(X_n,
Y_n)$, where $y_i = f(x_i)$, for $i=1,\dots,n=6$.  We privilege Gaussian
process (GP) based-surrogates \citep[e.g.,][]{gramacy2020surrogates}, but our
methodology is agnostic to that choice so long as the predictive equations
from $\hat{f}_n$ have similar features -- e.g., non-linear predictive mean
$\mu_n(x)$ and higher predictive uncertainty/standard deviation $\sigma_n(x)$
away from the training data sites.

\begin{figure}[ht!]
\centering
\includegraphics[scale=0.5,trim=0 0 0 25]{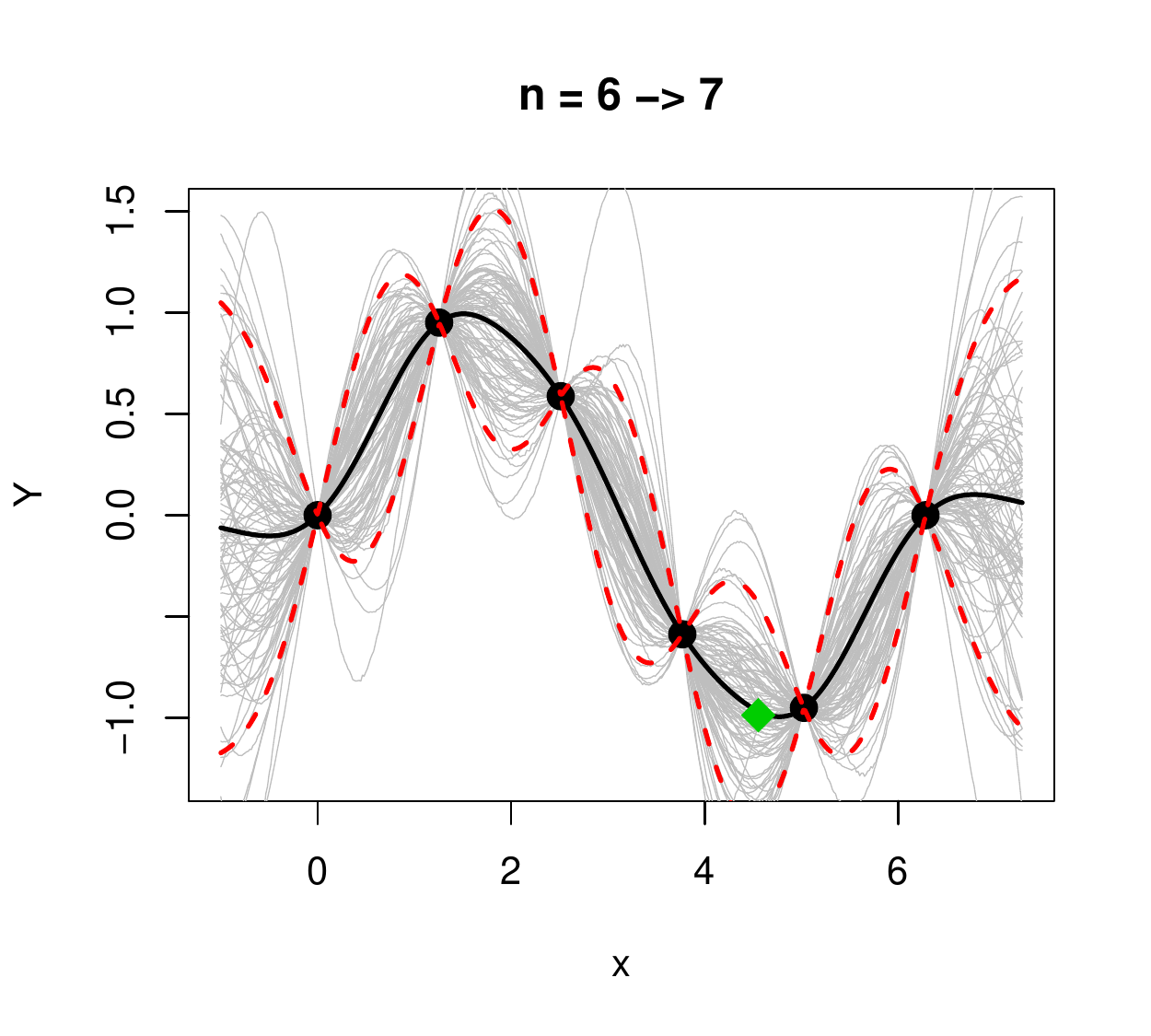}
\includegraphics[scale=0.5,trim=0 0 0 25]{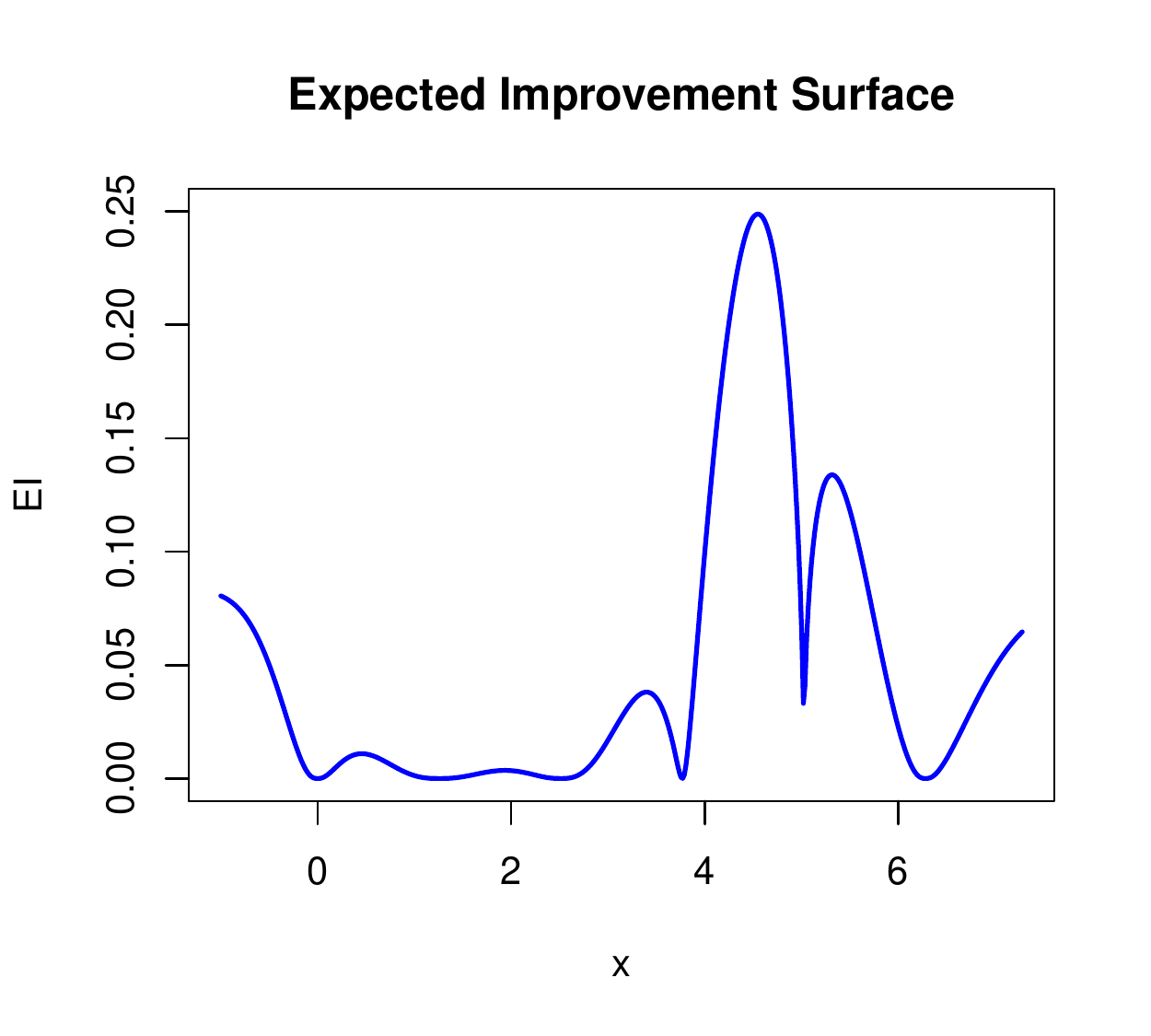}
\vspace{-0.25cm}
\caption{Predictive surface (left) via means (solid
black), 90\% intervals (dashed-red), and sample paths (gray); EI (right) and
resulting acquisition (green diamond).%, $n = 6\rightarrow 7$).
\label{fig:ei}}
\end{figure}

The left panel of Figure \ref{fig:ei} shows a fitted $\hat{f}_6$ via
$\mu_6$ as solid black line with error bars $\mu_6 \pm 2 \sigma_6$ in
dashed-red. Each of the one-hundred gray lines is a sample
from the predictive distribution.  This distribution interpolates
the training data (black dots), a GP hallmark, yet our contributions are
not limited to noise free settings. (We entertain a noisy $f$ in Section 
\ref{sec:ato}.)  
%Although we shall exclusively
%entertain $f$'s without noise, this is not a limitation of BO nor our own
%contributions.  
%Further discussion is tabled until Section \ref{sec:discuss}.
The current best run $f_n^{\min} = \min\{y_1,\dots,y_n\}$ is near $x =
5$, but there is considerable predictive uncertainty about other %, untried
parts of the input space $\mathcal{B}$, relative to $f_n^{\min}$.  
%Just eyeballing it, 
Locations in $[-1,0]
\cup [4,6] \cup [2\pi,2\pi+1]$ are promising.

BO {\em acquisition criteria} serve to
operationalize this notion.  We shall focus on two and note that
several others are variations on similar themes.  One is {\em Thompson
sampling} \citep[TS;][]{thompson1933likelihood}, and involves drawing from the
predictive distribution $Y(x) \sim \hat{f}_n \mid X_n, Y_n$ and choosing
$x_{n+1} = \mathrm{argmin}_{x\in
\mathcal{B}} \, Y(x)$.  TS amounts to randomly selecting one of those gray lines
and optimizing it in lieu of the expensive $f$, so the criterion is inherently
stochastic.  The other is {\em expected improvement}
\citep[EI;][]{jones1998efficient}.  Define {\em improvement} as $I(x) =
\min\{0, f_n^{\min} - Y(x)\}$ and take its expectation with respect to $Y(x)$
.  If  $Y(x) \sim \hat{f}_n$ is Gaussian, as with GPs, then this has
closed form:
\begin{align}
\mathrm{EI}(x) & = \mathbb{E}\{I(X)\} = \int I(x) \; dY(x) \label{eq:ei} %\\
%& 
= (f_n^{\min} - \mu_n(x))\,\Phi(z_n(x))
+ \sigma_n(x)\, \phi(z_n(x)), %\nonumber
\end{align}
where $z_n(x) = (f_n^{\min} - \mu_n(x))/\sigma_n(x)$ and $\Phi$/$\phi$
are the Gaussian cdf/pdf. 
For non-Gaussian $\hat{f}_n$, one can always resort to
Monte Carlo (MC) integration instead.  The right panel of Figure~\ref{fig:ei}
provides EI for $\hat{f}_6$. Contrary to TS, the EI acquisition
$x_{n+1} = \mathrm{argmax}_{x\in
\mathcal{B}} \, \mathrm{EI}(x)$ can be deterministic, if properly solved.
The maximum EI acquisition $x_7$ is shown as a green diamond on
the left panel. %The right column of panels in Figure \ref{fig:ei} shows the
%resulting surfaces after five such EI-optimizing acquisitions.

Observe that TS and EI involve ``inner-optimizations'' over a criterion, a
task that may be even more challenging than the original problem
(\ref{eq:mp}).  Each gray line is highly multi-modal, as is the EI surface.
Both have about as many local optima as there are training data points $n$,
whereas $f$ has only two local minima in $\mathcal{B}$. Speedy
evaluation of $\mu_n(x)$ and $\sigma_n(x)$ relative to $f(x)$ saves
us, but only partly.  We still desire good $x_{n+1}$ without exhaustive
search or cumbersome subroutines.

Two strategies are common, sometimes separately, sometimes in tandem as a
hybrid.  The simplest option is to distribute $N$ candidate points
$\mathcal{X}_N$ throughout the input space, evaluate the criterion on
$\mathcal{X}_N$, and thereby replace a continuous search with a discrete one.
In low input dimension a dense grid of candidates can be used effectively.  In
higher dimension one can populate $\mathcal{X}_N$ with a space-filling design
like a random Latin hypercube sample \citep[LHS;][]{Mckay:1979} to manage the
computational expense of evaluating the criterion exhaustively.  A
higher-powered approach is to locally apply a smooth, convex
optimization library such as L-BFGS-B \citep{BFGS}.  Derivatives may be
approximated by finite-differencing or autograd \citep{paszke2017automatic}, or
may have a simple closed form depending on the criterion and nature of
$\hat{f}_n$. The former requires more computer work and some consideration of
numerical stability; the latter more researcher/programmer effort when
possible. (MC-based $\hat{f}_n$ or EI challenges both approaches.) Pure
candidate-based search is most common with TS because of its stochastic
nature. Gradient-based continuous search is popular in simple GP/EI setups,
but a multi-start scheme is essential to avoid inferior local solutions. This
is where the hybrids come in: candidates seeding local solvers.

In this paper we contend that both strategies, random candidates
and multi-start local optimization, can be replaced by (or hybridized with)
more thoughtfully chosen $\mathcal{X}_N$.  In the 1d setting of Figure
\ref{fig:ei} we could place candidates at the midway points between each of
the existing $n$ inputs and the boundary $\mathcal{B}$, implementing a kind of
multi-pronged bisection search \citep[][Section 2.1]{burden1985numerical} and
resulting in $N = n+1=7$ candidates.  The best of those $\mathcal{X}_N$ by
either criterion may not give a precise solution to the inner optimization,
but it would be an effective one because EI and many of the random gray lines
indicate a solution close to one of those midway points. Such locations might
be much better than ones identified by a limited candidate or numerical local
search.

This 1d example is overly simplistic.  Going forward, we shall explicitly
target two and higher dimensions.  In Section \ref{sec:tri} we scale-up
the midway candidate idea, also suggested by \citet{scott2011correlated}, to
what we call ``tricands'', based on Delaunay triangulation and the convex hull
of $X_n$.  We explore tricands' features and limitations and suggest remedies
with the BO application in mind. In Section \ref{sec:classic} we demonstrate
that tricands outperform both random LHS candidates and a multi-start
gradient-based numerical inner optimization in the conventional setting where
surrogate GP predictive equations are available in closed form. In Section
\ref{sec:nonstat} we consider two nonstationary surrogates requiring
Markov chain Monte Carlo (MCMC), for which closed-form acquisition criteria
are not readily available. Candidates are essential in this setting, and our
tricands are better than random space-filling ones. Our discussion in Section
\ref{sec:discuss} emphasizes tricands' ``plug-n-play'' nature -- they can
be inserted into any candidate-based scheme -- and suggests potential for
extension.%Perhaps the single
%greatest selling point for tricands is how easy they are to insert into an
%existing BO setup by swapping our implementation in place of another
%candidate-based scheme.

\section{Delaunay triangulation candidates}
\label{sec:tri}

Many criteria for BO resemble Eq.~(\ref{eq:ei}), balancing exploitation
($\mu_n(x)$ below $f_n^{\min}$) with exploration (large $\sigma_n(x)$). Most
surrogates inflate predictive uncertainty ($\sigma_n(x)$) away from training
data locations $X_n$.  In the case of GPs, this is what produces the ``sausage
shaped'' predictive intervals shown as red-dashed lines in Figure
\ref{fig:ei}. Our main insight is that careful allocation of candidates {\em
between} existing training data locations, where $\sigma_n(x)$ is high, allows
for BO acquisitions that do not necessitate cumbersome numerical optimization
of posterior predictive quantities but still balance exploitation and
exploration. A hard statistical optimization can be replaced with an easier
geometric one.

\begin{figure}[ht!]
\centering
\includegraphics[scale=0.54,trim=0 0 0 58,clip=TRUE]{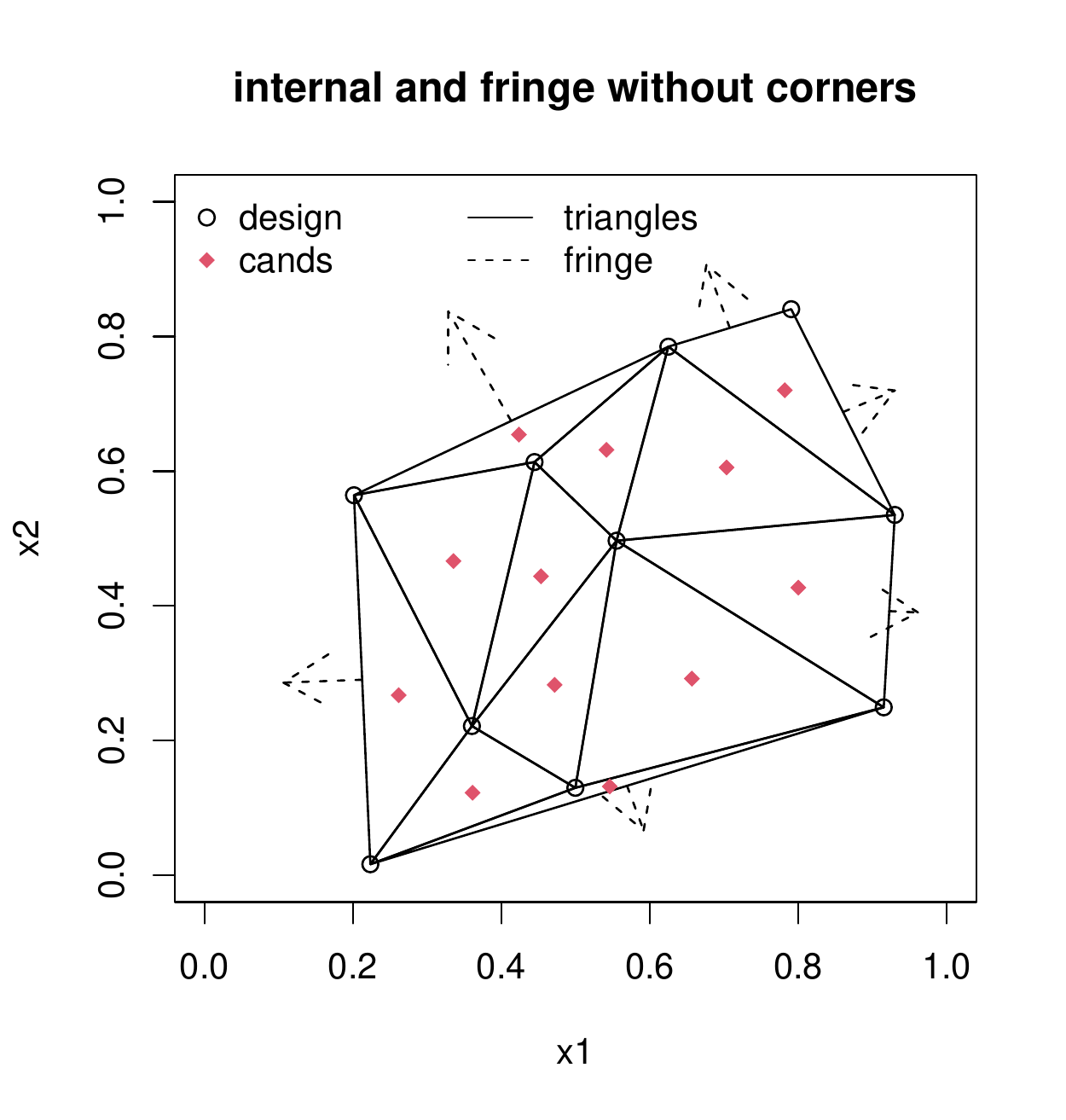} 
\includegraphics[scale=0.54,trim=0 0 0 55,clip=TRUE]{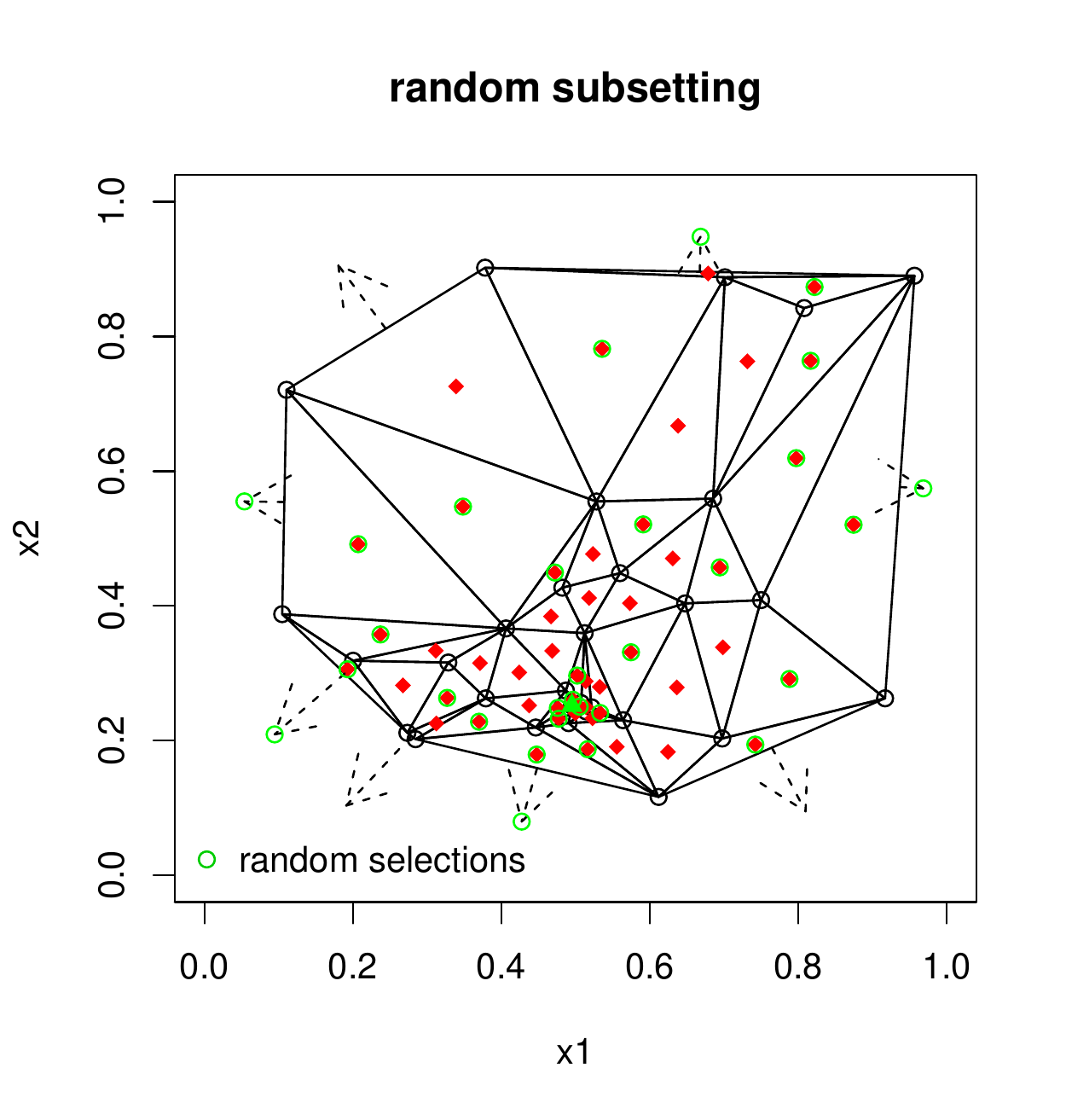}
\vspace{-0.25cm}
\caption{Interior and fringe candidates (both), and randomly sub-sampled candidates (right).
\label{fig:tricands}}
\end{figure}

The idea is sketched graphically in Figure \ref{fig:tricands}, whose details
will emerge over the next three subsections.  The existing design $X_n$ is
shown as open circles, and the selected candidates are closed red dots {\em
and} the tips of dashed arrows.  We call these interior (Section
\ref{sec:interior}) and fringe (Section \ref{sec:fringe}) candidates,
respectively.  Both are calculated based on Delaunay triangles and their
convex hull, outlined by solid black lines in the figure, and explained next.
We provide this illustration in 2d to ease visualization; however our
Section \ref{sec:classic}--\ref{sec:nonstat} benchmarks include higher
dimensional input spaces.

\subsection{Interior candidates}
\label{sec:interior}

A {\em Delaunay triangulation} of $X_n$ is an angle-maximizing set of $d$
line-segments connecting geographically nearby points, such that no point lies
inside the circumcircle of those points. Such triangulations subdivide the
interior of the {\em convex hull} of $X_n$, which is the the smallest (convex)
set that contains all points.  In 2d, lines depicting those subdivisions
form triangles.  In higher dimension they form tetrahedra, however one often
abuses the nomenclature and still refers to triangles. For more details, see,
e.g., \citet{lee1980two}.  
%
% Delaunay triangulation is the dual of a Voronoi
% tesselation, which has been deployed for divide-and-conquer-based
% nonstationary surrogate modeling
% \citep[e.g.,][]{kim2005analyzing,rushdi2017vps}, imposing both statistical and
% computational independence.  
%
The solid lines of Figure \ref{fig:tricands} indicate a triangulation for
random $X_n$ (left) and an $X_n$ derived after BO iterations (right),
described in Section \ref{sec:classic}
.

There are fast algorithms for calculating Delaunay triangulations. In 2d,
these have $\mathcal{O}(n \log n)$ runtimes.  Higher dimensional analysis is
complicated by the number of triangles, which depends on the geometry of
$X_n$, a topic we shall return to shortly. For {\sf R} we use the {\tt
geometry} package on CRAN \citep{geometry}; for {\sf Python} we use {\tt
Delaunay} in {\tt scipy.spatial} \citep{2020SciPy-NMeth}.  Both are wrappers
around the {\sf C} library {\tt Qhull}, implementing ``quickhull''
\citep{quickhull}. \citet{bates2001tri} first suggested Delaunay triangulation
for BO. That early work only explored two input dimensions, possibly because
they did not have convenient access to {\tt Qhull}.  They also did not
entertain the BO-specific extensions we provide here, particularly in Sections
\ref{sec:fringe}--\ref{sec:subset}.

Let the triangles be denoted by $T_j$, for $j=1,\dots,n_T$. 
% Allow us to table a discussion of how $n_T$ relates to the design size $n$. 
Each $T_j$ is a $(d+1) \times d$ matrix when $X_n$ has $d$ columns. 
% In 2d this
% is $3 \times 2$, providing the vertices of a triangle.  In higher dimension
% these are the vertices of a tetrahedron. 
Create $n_T$ new candidates
$\mathcal{X}_{n_T}$ where the $j^\mathrm{th}$ candidate is located at the {\em
barycenter} of $T_j$:
$$
\tilde{x}_j = \bar{T}_j = \frac{1}{d+1} \sum_{i=1}^{d+1} T_j[i,]
\quad \mbox{or} \quad \tilde{x}_{jk} = \frac{1}{d+1} \sum_{i=1}^{d+1} T_j[i,k],
\;\;\; \mbox{for } k=1,\dots,d.
$$
The left expression is vectorized over the second, column dimension of $T_j$.
The second is explicit about  coordinates $\tilde{x}_j^\top = (\tilde{x}_{j1},
\dots, \tilde{x}_{jd})$.  Red dots in Figure \ref{fig:tricands} provide a
visual. These locations will almost certainly not be the maximal points of
$\sigma_n(x)$ in the vicinity of $T_j$, but they will be close because
$\tilde{x}_j$ is within $T_j$ but far from its edges. We refer to these
$\mathcal{X}_{n_T}$ as ``interior'' candidates.

In 2d, Euler's formula gives that $n_T = 2n - 2 - h(X_n)$ where $h(X_n)$ is
the number of elements of $X_n$ on its convex hull.  In Figure
\ref{fig:tricands}, $n=10$ and $h(X_n) = 6$ so $n_T = 12$.  When $d \geq 3$,
the number of faces of the tetrahedra can grow as $n^{\lceil d/2 \rceil}$
depending on the nature of $X_n$.
\begin{figure}[ht!]
\centering
\includegraphics[scale=0.45,trim=0 0 0 59,clip=TRUE]{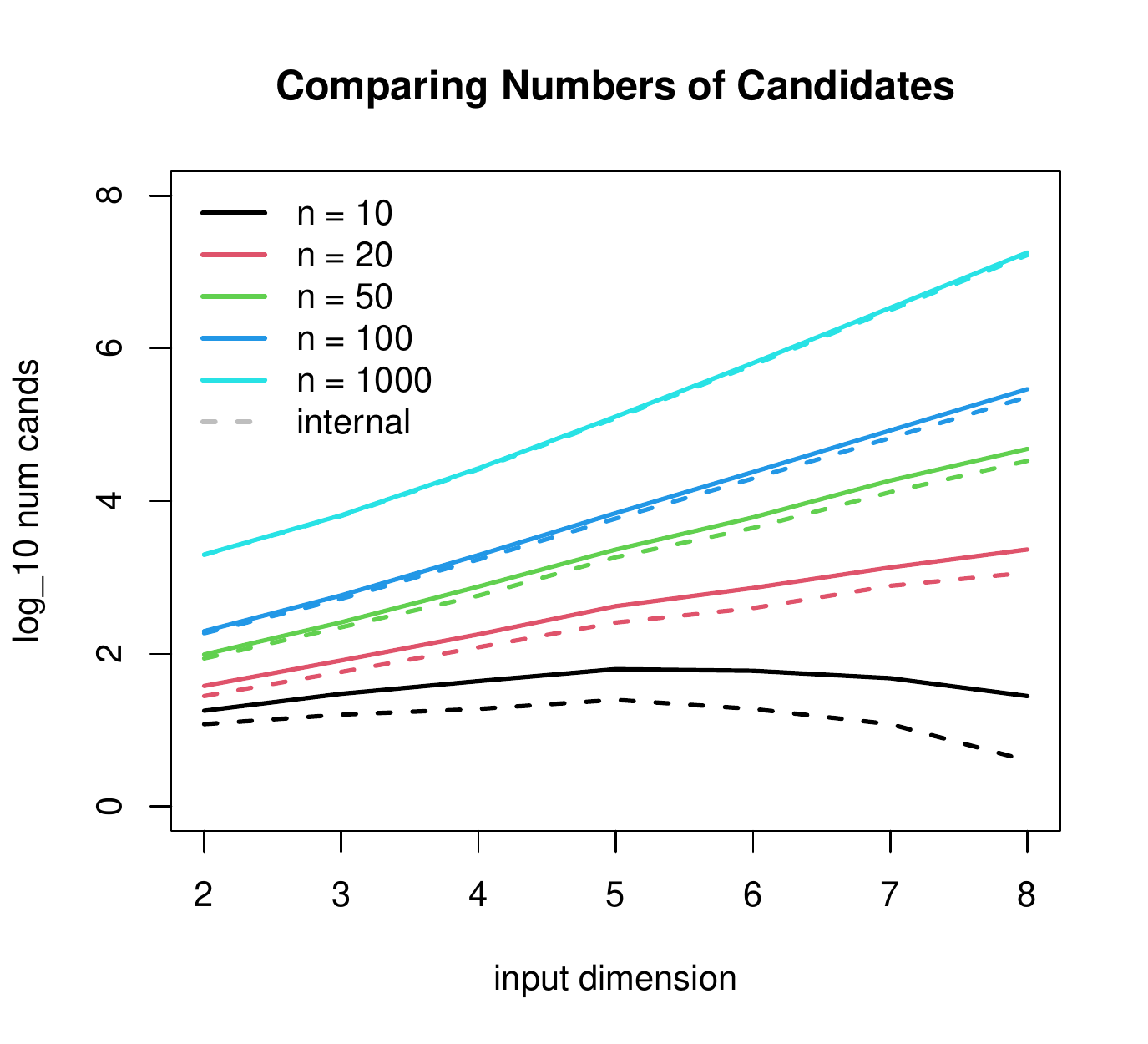}
\includegraphics[scale=0.45,trim=0 0 0 59,clip=TRUE]{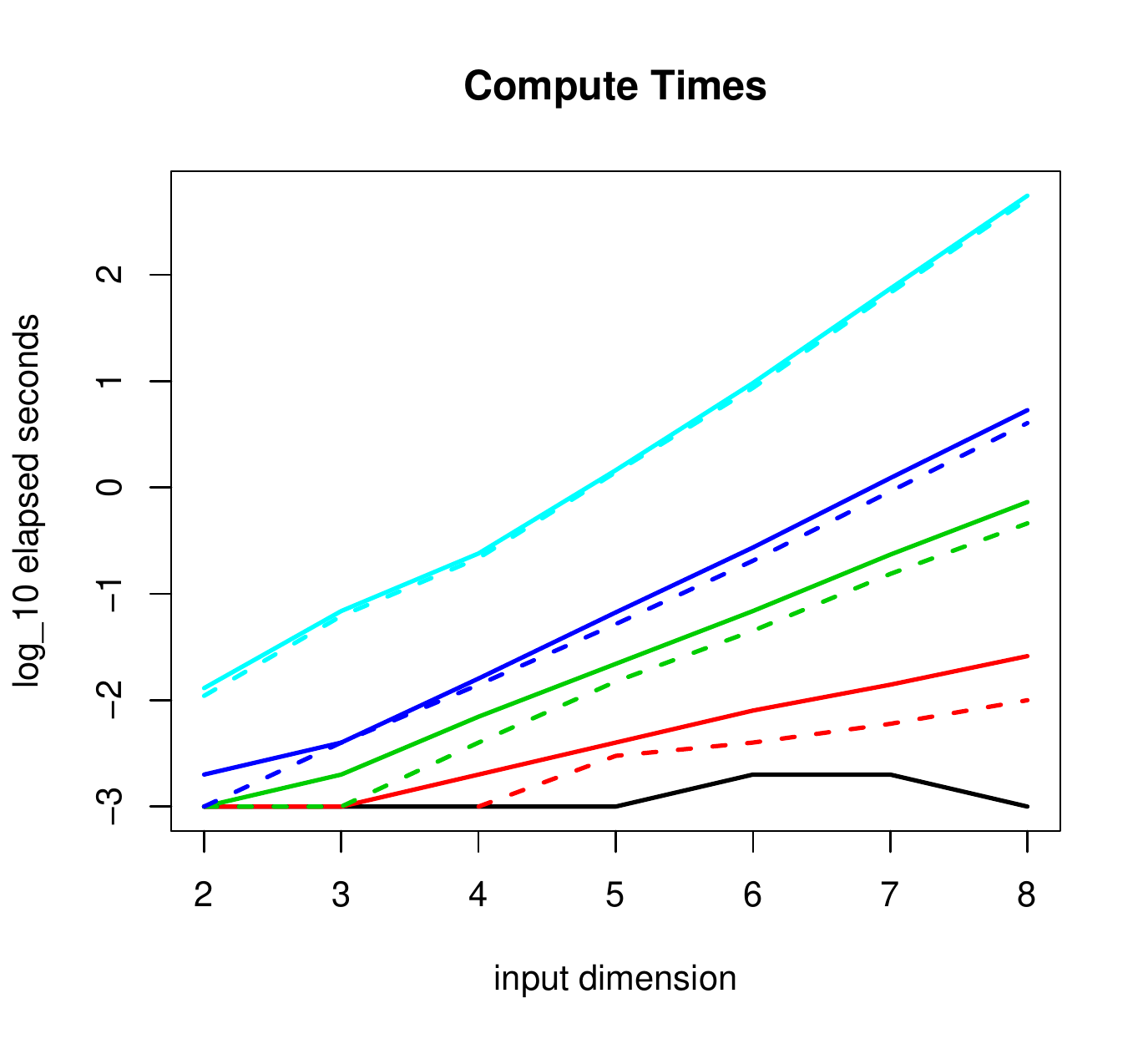}
\vspace{-0.25cm}
\caption{Tricands $N$ (left) over input dimension $d$ and design size $n$, and
compute time (right) using a single core of an Intel i7-6900K CPU at 3.20GHz.
\label{fig:size} }
\end{figure}
Figure \ref{fig:size} provides an empirical view of the number of candidates
(left) and triangulation compute time (right), for varying $d$ and $n$ where
$X_n$ is distributed uniformly at random in $\mathcal{B} = [0,1]^d$.  The
dashed lines in the left panel are $\log_{10} n_T$.  Notice that when
$n$ is small, $n_T$ decreases as $d$ increases.  For fixed $d$, $n_T$ steadily
increases with $n$.  In our BO context this is a good thing, because so too
does the modality of acquisition criterion.  With nonparametric surrogates
like GPs, the complexity of the response surface can increase with training
data size $n$. The search effort for the next acquisition should be
commensurate with this complexity, an innate characteristic of (interior)
tricands. The right panel of Figure \ref{fig:tricands}, described later in
Section \ref{sec:subset}, shows how this works after many iterations of EI
acquisition.

\subsection{Fringe candidates}
\label{sec:fringe}

Having candidates $\mathcal{X}_{n_T}$ only in the interior of the convex hull
could limit exploration if the complement of the volume of the hull and
$\mathcal{B}$ is large.  It could limit exploitation if the solution is
on/near the boundary of $\mathcal{B}$.  One remedy is to force $X_n$ to contain
boundary points, such as the corners of the input space. This is not a bad
approach in low dimension (e.g., in 2d there are only four corners), but
could be prohibitive in higher dimension.  Later in Section \ref{sec:classic}
we consider a $d=8$ example, having 256 corners in $\mathcal{B}=[0,1]^8$, which
would nearly blow our entire budget of runs.  

We instead prefer the ``fringe'' candidates pointed to by the dashed arrows in
Figure \ref{fig:tricands}. There is one of these for each facet (edge in 2d)
of the convex hull, extending perpendicularly from middle of the facet half
way to the boundary of $\mathcal{B} = [0,1]^d$.  The {\tt Qhull} library
furnishes both ingredients: $d
\times d$ facets $F_j$ and normal $d$-vectors $\vec{v}_j$ for $j=1,\dots,n_F$.
Using these quantities, let $\bar{F}_j = \frac{1}{d} \sum_{i=1}^d F_j[j,]$
denote the coordinates of the middle of facet $j$.  Then, the $j^\mathrm{th}$
fringe candidate in $\mathcal{X}_{n_F}$ is
\[
\vec{x}_j = \bar{F}_j + {\textstyle \frac{1}{2}} \alpha_j \vec{v}_j \;\;\; \mbox{where} \;\;\;
\alpha_j = \min \left\{ \mathbb{I}_{\{\vec{v}_j > 0\}} - \bar{F}_j \vec{v}_j \right \}.
\]
Above, ``$\min$'' is picking out the nearest boundary to $\bar{F}_j$. Division
by two mirrors the bisection search analogy for interior candidates, but this
could be a tuning parameter.  Non-unit rectangular $\mathcal{B}$ is doable,
but requires a more convoluted formula.  More general $\mathcal{B}$ may
present challenges.

Fringe and internal candidates may be combined: $\mathcal{X}_N =
[\mathcal{X}_{n_T}; \mathcal{X}_{n_F}]$, stacked row-wise to form an $N \times
d$ matrix with $N=n_T + n_F$.  The number of fringe candidates $n_F$ is
generally small compared to $n_T$.  This is shown empirically in the left
panel of Figure \ref{fig:size}, where $N$ is indicated by the solid line, and
$n_T$ by the dashed one.  As $n$ increases, the gap between $N$ and $n_T$,
indicating $n_F$, narrows. % until it is not visible on the log scale.

\subsection{Targeted sub-sampling}
\label{sec:subset}

Having $N$ grow exponentially in $n$ may not suit all applications.
Entertaining $N \approx 10{,}000$ candidates when $n=100$ and $d=6$, referring to
Figure \ref{fig:size}, is cumbersome and potentially overkill. One way to
limit $N$ is random sub-sampling: simply calculate the full $\mathcal{X}_N$
based on $X_n$ and downsample, uniformly at random, a subset of size
$N_\mathrm{sub} < N$.  Since our triangulation strategy was designed to
focus on exploration (finding locally high $\sigma_n(x)$) via midway
candidates, we find it advantageous to guarantee retaining some of those
candidates which are promising for exploitation (low $\mu_n(x)$), still
without any explicit numerical optimization -- only geometry.

% \begin{figure}[ht!]
% \centering
% \includegraphics[scale=0.6,trim=0 0 0 55,clip=TRUE]{rand1}
% %\includegraphics[scale=0.64,trim=0 0 0 50,clip=TRUE]{rand2} 
% \vspace{-0.5cm}
% \caption{Illustrating interior and fringe candidates with the 
% Goldstein--Price problem (Section \ref{sec:gp}) at $n=30$.  Legend
% details similar to Figure \ref{fig:tricands}; randomly subsampled candidates
% are circled in green.  
% \label{fig:subset} }
% \end{figure}

One of the rows of $X_n$ corresponds to $f_n^{\min}$, the best input found so
far: $x_n^{\min} = x_i$ s.t.~$i =
\mathrm{argmin}_{i=1,\dots,n} \, y_i$ (or $\mu_n(x_i)$ in the noisy case).
Let $\mathcal{T}_n^{\min} = \{T_j : x_n^{\min} \in T_j, j=1,\dots,n_T\}$
denote the set of triangles containing $x_n^{\min}$, and similarly let
$\mathcal{X}_n^{\min} = \{ \tilde{x}_j \in
\mathcal{X}_{N} : T_j \in \mathcal{T}_n^{\min} \}$ denote the candidates 
 associated with those triangles.  Those are, in a geometric sense,
adjacent to $x_n^{\min}$.  A fringe candidate may also be considered
adjacent in an analogous way, however we place them in the
complement $\mathcal{X}_N \setminus \mathcal{X}_n^{\min}$.  Now, rather than
sub-sample uniformly from the full set $\mathcal{X}_N$, we partition sampling
from points adjacent to $x_n^{\min}$, i.e., from $\mathcal{X}_n^{\min}$, and
from points farther afield in $\mathcal{X}_N
\setminus \mathcal{X}_n^{\min}$.  In so doing, we guarantee that our
$N_{\mathrm{sub}}$ candidates cover potential for exploitation and
exploration, respectively. We prefer a 10:90 split, with up to 10\% of
$N_{\mathrm{sub}}$ coming from candidates adjacent to $x_n^{\min}$, fewer if
$|\mathcal{X}_n^{\min}| < N_{\mathrm{sub}}$, and likewise 90\% from its
complement.  %Additional tuning could be advantageous.

The right panel of Figure \ref{fig:tricands} provides an illustration after
$n=30$ runs optimizing the Goldstein--Price function with EI under random
initialization (details in Section \ref{sec:gp}).  When $n=30$ we have $N
  \approx 60$, with the precise value depending on $h(X_n)$. Here we consider
$N_{\mathrm{sub}} = 30$. Observe how randomly sub-sampled candidates
$\mathcal{X}_{N_{\mathrm{sub}}}$, circling the original red candidates
$\mathcal{X}_N$ in green, concentrate near the global minimum $(0.5, 0.25)$
because $x_n^{\min}$ is in the vicinity. Other sub-sampled candidates are
spread out more widely.  This figure illustrates how acquisitions, and thus
candidates, gravitate toward promising regions for exploitation without
neglecting areas of potential exploration.  A ridge of local minima may be
found in the banana-shaped region traced out by a concentration of $X_n$ and
$\mathcal{X}_N$. % in both views.

\subsection{Implementation and software}

Our implementation, provided in {\sf Python} and {\sf R} in our git
repository,\footnote{\url{http://bitbucket.org/gramacylab/tricands}} 
% \footnote{\url{http://bitbucket.org/blinded/tricands} (see zip file in supplementary material)} % UNBLINDED
is relatively tidy.  For example, {\tt tricands.R} therein contains just 71
lines of code, eleven of which are to support optional visualizations in 2d
such as those in Figure \ref{fig:tricands}.  The heavy lifting is done by {\tt
Qhull}. When performing Delaunay triangulations, we provide option
\verb!"Q12"! to  work around numerical instabilities that sometimes arise.
When calculating convex hulls, option \verb!"n"! returns normal vectors used
in calculating fringe candidates (Section \ref{sec:fringe}).

Defaults yield both fringe and internal candidates and fix a maximum candidate
size of {\tt max} $ = N_{\mathrm{sub}} = 100d$, but these are user-adjustable.
In experiments coming shortly, we deliberately limit $N_{\mathrm{sub}}$ even
further so that a fairer comparison can be made to other continuous search and
candidate-based methods. When $N\leq N_{\mathrm{sub}}$, all internal and
fringe candidates are returned. The user may supplement these with additional
$N_{\mathrm{sub}} - N$ random candidates if desired. If $N_{\mathrm{sub}} <
N$, the execution flow looks for a variable {\tt best}, providing the index of
$x_n^{\min}$ in $X_n$, making sure about 10\% of tricands come from adjacent
triangles. When
\verb!best=NULL!, tricands are sub-sampled at random.

\section{Classical GP benchmarking}
\label{sec:classic}

In this first of two sections on benchmarking, we focus on BO via traditional
GP surrogates. We follow a homework problem in Section 7.4 of
\citet{gramacy2020surrogates}, which piggy-backs off of GP, EI, and TS
demonstrations earlier in the chapter.  Software and other particulars are
relegated to Appendix \ref{app:implement}.  All of our examples
are fully reproducible using the code provided in our git repository.
We consider three methods for solving EI acquisitions: a continuous search of
the criterion via L-BFGS-B with 5-multi-starts, LHS candidates, and tricands.
For TS acquisitions, we similarly employ both LHS candidates and tricands. As
non-BO (not surrogate-based) comparators, serving primarily as benchmarks,
we entertain ``raw'' L-BFGS-B and Nelder--Mead \citep{nelder1965simplex}.
%\footnote{Our experimental apparatus additionally included probability
%of improvement (PI) and upper-confidence-bound (UCB) acquisitions. However we
%dropped them from this presentation for clarity because PI was uniformly
%dominated by EI, and UCB matched EI for reasonable values of the tuning
%parameter.}

 % These under-perform the BO alternatives, but nevertheless serve as an
% important benchmark.

Two examples are showcased here, with an third example relegated to
Appendix \ref{app:hart6}. Our synthetic $f$s, including those in Section
\ref{sec:nonstat}, are described in more detail on the pages of the Virtual
Library for Simulation Experiments \citep[VLSE;][]{surjanovic2013virtual}.  In
all of our experiments we code inputs to $[0,1]^d$ and summarize results for
100 random restarts where each surrogate is initialized with the
same (unique to each random restart) starting design of size $n_0 = 12$,
except in Section \ref{sec:ato} where we use $n_0 = 60$.  This design is taken
uniformly at random following the advice of \citet{zhang2021distance} who
caution that small space-filling initial designs can spark pathological
behavior in BO.  We track ``best observed value'' (BOV) $f_n^{\min}$ as a
measure of progress which is summarized by median over
$n=1,\dots,n_{\mathrm{end}}$ and by boxplots for particular $n$ along the way.

% We also control and keep track of the number of evaluations of the acquisition
% criteria for each method. Other problem-specific details are provided as
% needed.
%
% To foreshadow somewhat, these two examples are chosen because they work well
% with a traditional GP/EI setup. In particular, the deployment of standard,
% library-based GPs means that criteria like EI can be optimized with library
% methods.  This is in contrast to the examples coming next in Section
% \ref{sec:nonstat}. Those are not well-fit by ordinary GPs because of their
% regime-changing behavior, and so we must entertain higher-powered alternatives
% with more limited support for solving for acquisitions. 

\subsection{Goldstein--Price}
\label{sec:gp}

\begin{figure*}[ht!]
\centering
\includegraphics[scale=0.44,trim=5 10 25 10,clip=TRUE]{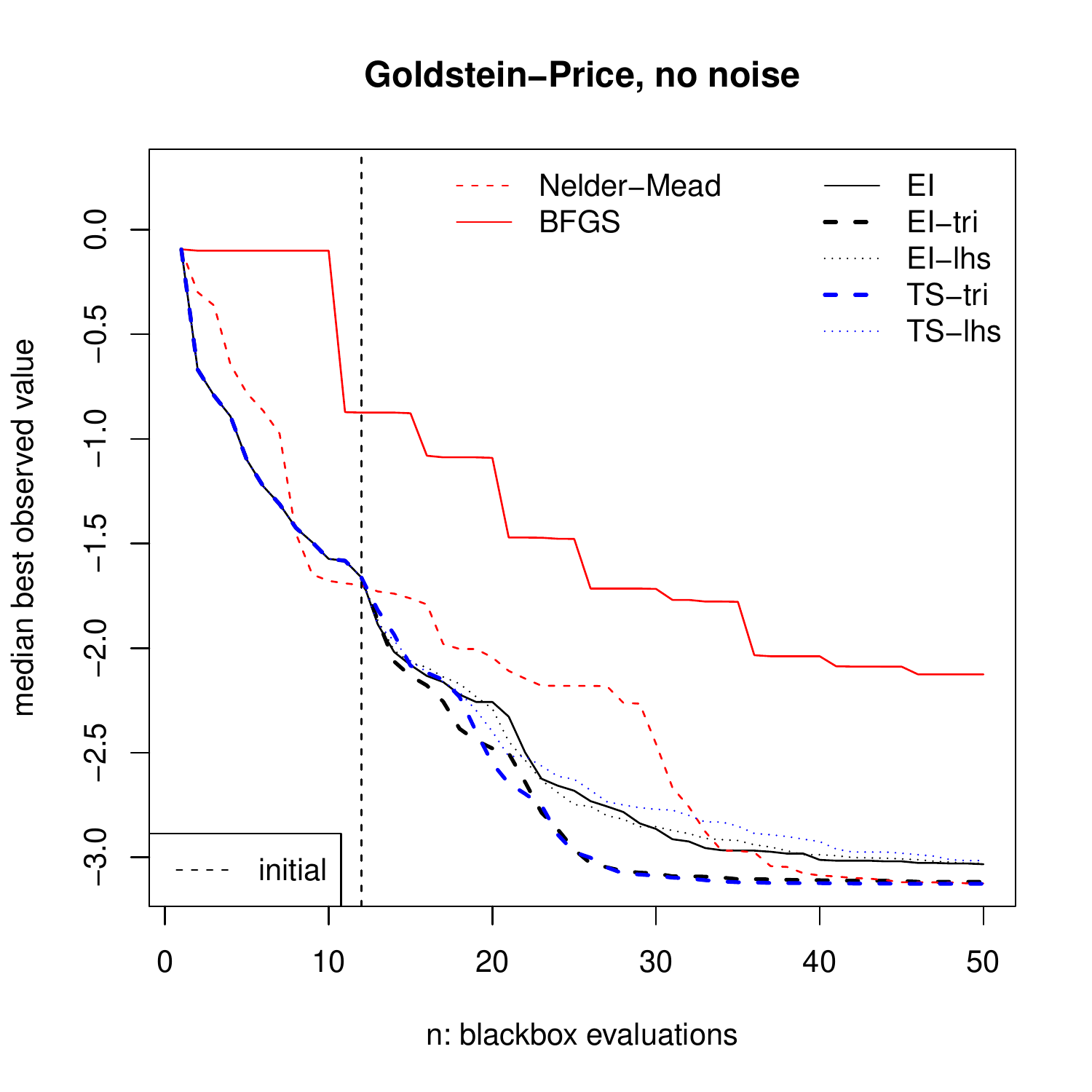}
\includegraphics[scale=0.44,trim=5 10 25 10,clip=TRUE]{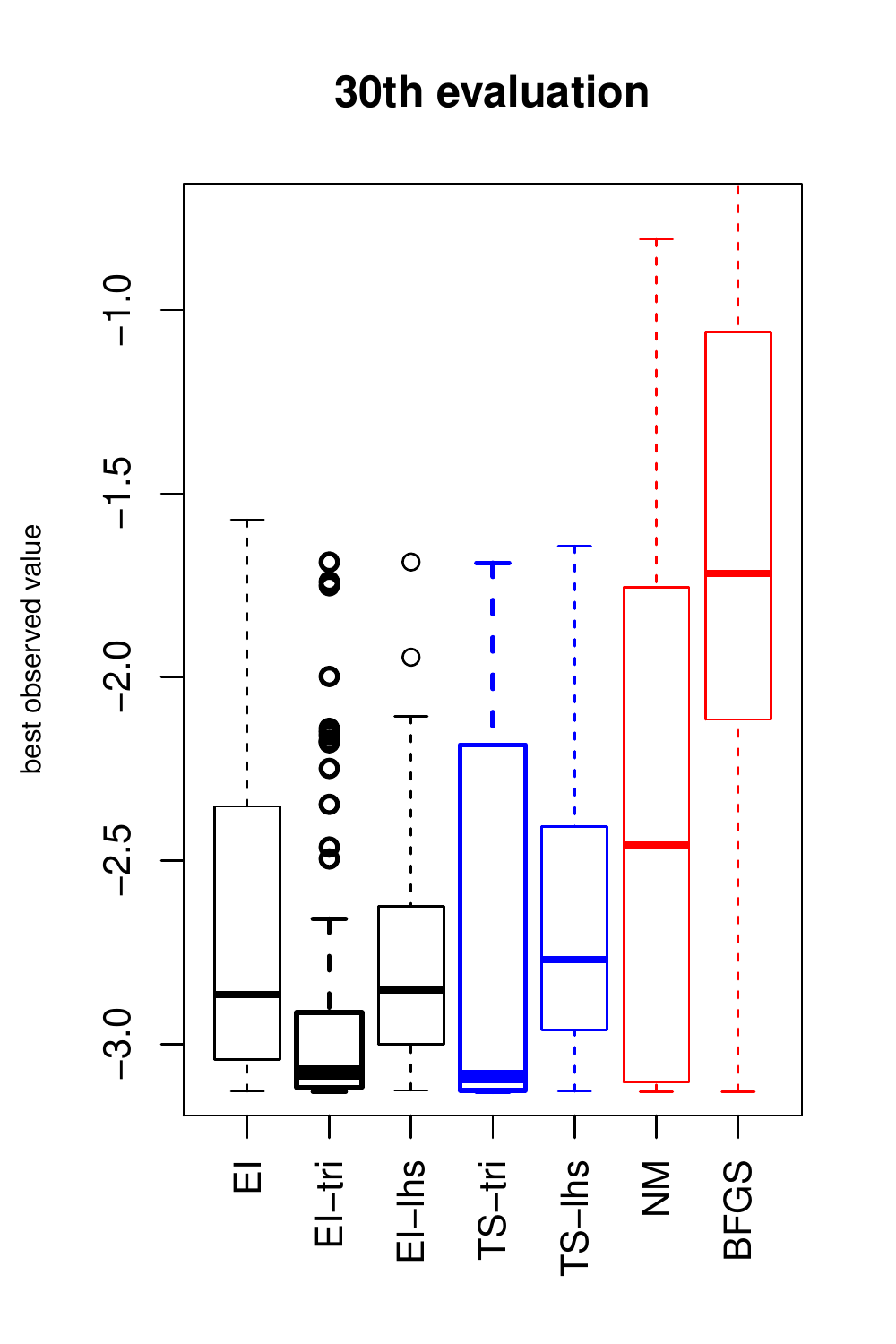} 
\includegraphics[scale=0.44,trim=50 10 20 10,clip=TRUE]{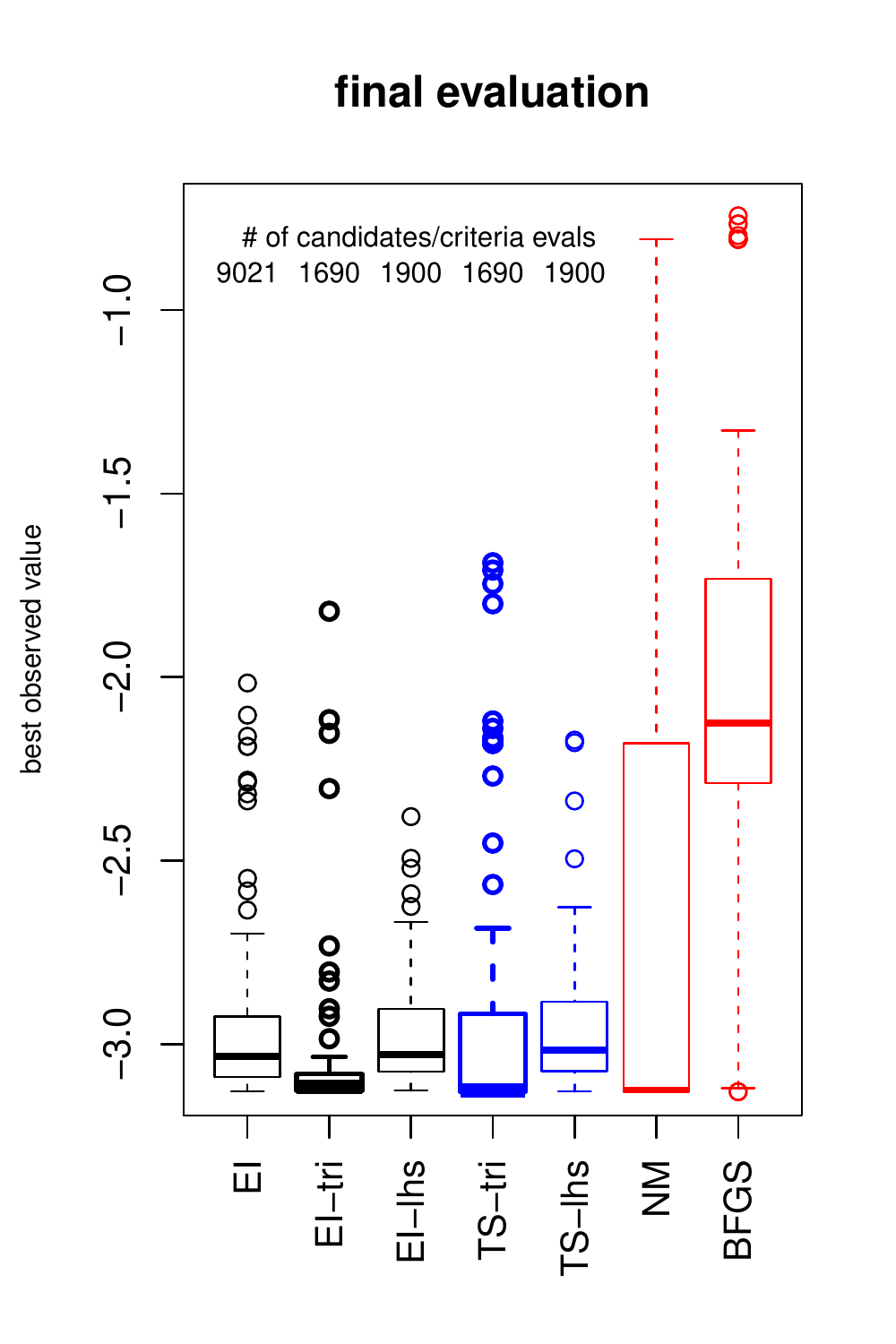} 
\includegraphics[scale=0.44,trim=5 10 25 10,clip=TRUE]{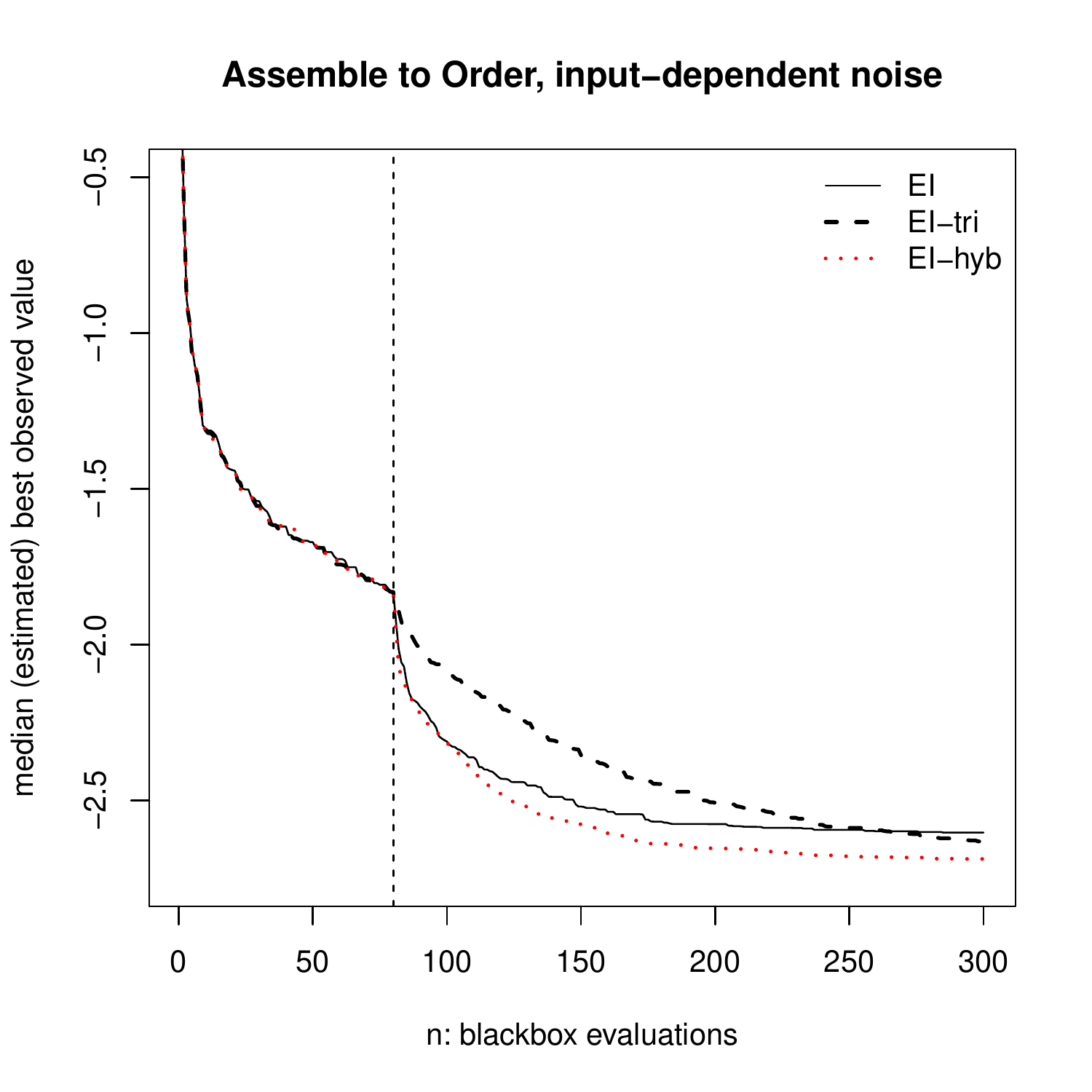}
\includegraphics[scale=0.44,trim=5 10 25 10,clip=TRUE]{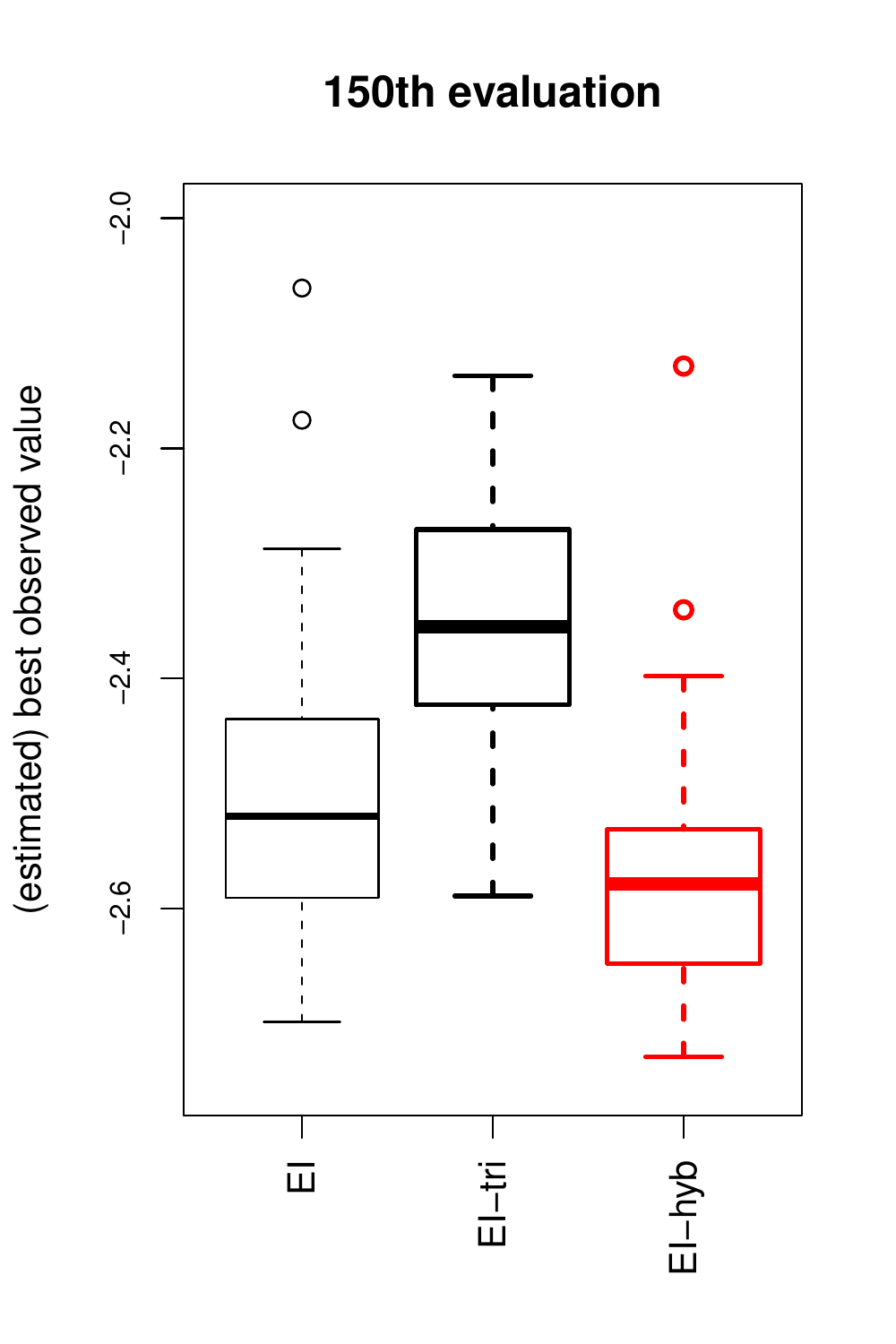} 
\includegraphics[scale=0.44,trim=50 10 20 10,clip=TRUE]{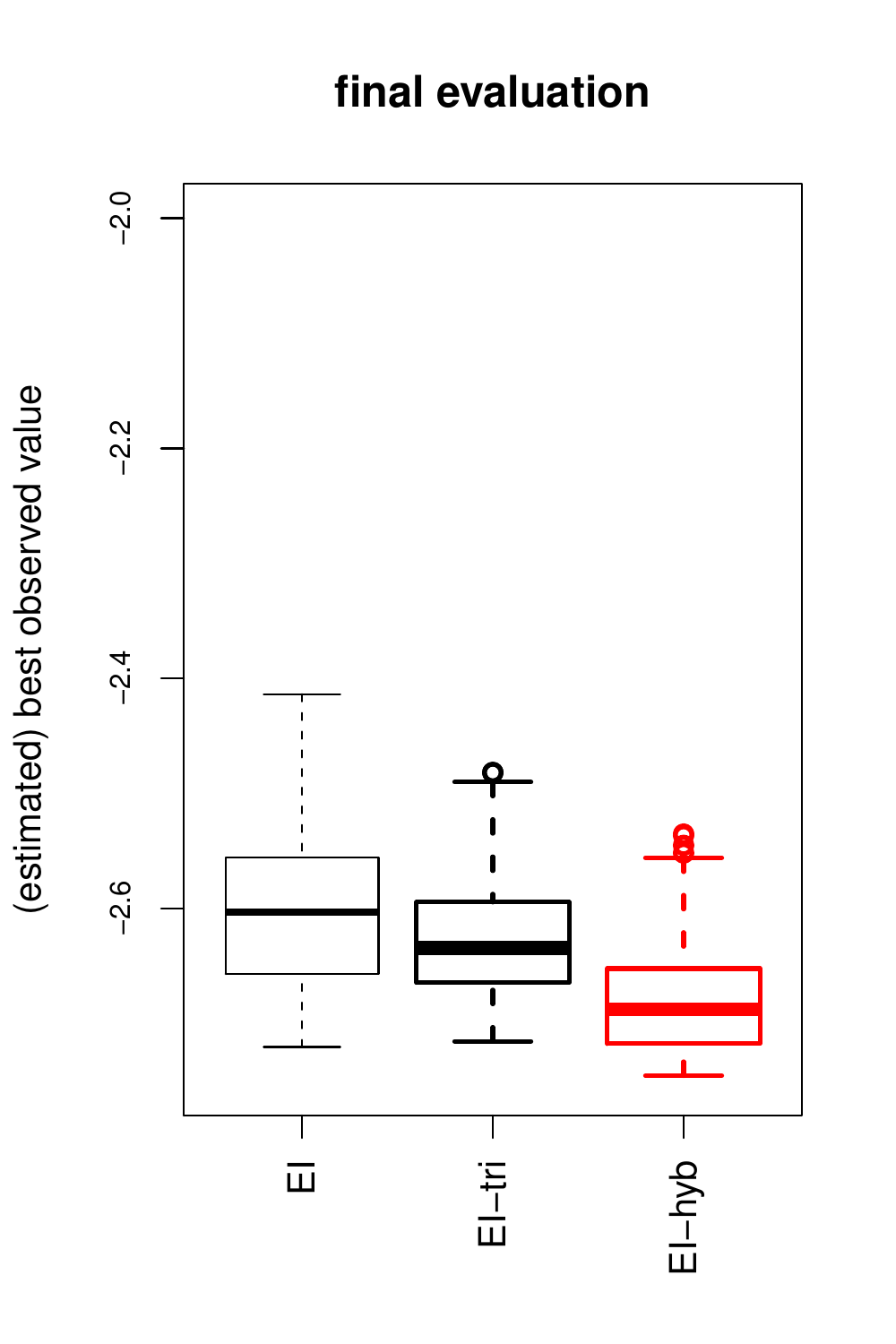} 
\vspace{-0.25cm}
\caption{Goldstein--Price (top) and ATO (bottom) BOV over 100 MC
trials: median (left panel) and distribution at intermediate (middle) and final (right) 
acquisitions.  Suffixes ``-tri'' or ``-lhs'' denote a candidate search.
Absence of a suffix (e.g., ``EI'' alone) indicates multi-start
L-BFGS-B.
\label{fig:gphart}}
\end{figure*}

The Goldstein--Price
function
%\footnote{\url{http://www.sfu.ca/~ssurjano/goldpr.html}} 
is a popular low-dimensional (2d) benchmark for BO
\citep[e.g.,][]{picheny2012benchmark}. A total of $n_{\mathrm{end}} = 50$
acquisitions are entertained, and all candidate-based methods (tricands and
LHS) are limited to fifty candidates. This means {\tt max} $= N_{\mathrm{sub}}
= 50$ for tricands, with fewer candidates when $N <
N_{\mathrm{sub}}$.  The top row of Figure \ref{fig:gphart} summarizes results
in three views.  In the top-left panel, median progress in BOV is shown over
increasing budgets of evaluations ($n$) as if each subsequent acquisition were
the last.  In the top-middle and right panels, boxplots capture the
distribution of BOV at $n=30$ and $n=50$, respectively.  In the top-left
panel, methods based on a multi-start numerical local search use solid lines;
those based on tricands are dashed; those based on LHS candidates are dotted.
(Nelder--Mead is an exception, being red-dashed.)  In the boxplots, tricands
use slightly heavier ink so that they stand out. Text printed at
the top of the top-right panel indicates the average number of times the
criterion, EI or TS, was evaluated in solving the inner optimization
sub-problem(s), cumulative over all acquisitions.  In the case of a
numerically optimized EI (labeled ``EI''), these happen within iterations of
L-BFGS-B search.

Beginning with medians over $n$ in the top-left panel, observe that the dashed
lines (those based on tricands) are uniformly superior to all other
comparators.  EI with tricands is slightly better than TS with tricands.  Only
at the very end of the run does the raw Nelder--Mead alternative look competitive.
At the 30th evaluation (top-middle) the boxplots indicate that there are
some MC repetitions where BOV based on tricands are underperforming.
However, more than half of the time these are the best two (i.e., both EI and
TS with tricands) of all.  TS with tricands is the winner in terms of
best-case performance, whereas the EI version has slightly better worst-case
behavior. Neither local optimizer (red) is competitive.  Finally, in the
top-right panel we see that the story is similar, except that now EI with tricands
edges out its TS analog.  Although median performance for Nelder--Mead
is competitive, more than half of the time its BOV at $n_{\mathrm{end}} = 50$
is among the two worst in the experiment.  

% All of the methods have some relatively high outlying BOV values, although we
% caution that the scale of the $y$-axis of the middle and right plots is
% different than that for the top-left panel, showing only median progress.  As
% always with BO, you can get unlucky with initialization, or conversely lucky
% with other random aspects of a search. This explains many of the outlying dots
% in the boxplots for all BO comparators, none of which is consistently
% under-performing.  All find the answer with low variability using a
% slightly larger $n_{\mathrm{end}}$, 
% but extending out in that direction makes
%  for a less interesting comparison.% \added{(just a note: this description is good, but if we need to cut down on the 
% length of the paper this may be a good place to do it.)}

Perhaps the most striking result from the figure is that tricands-based EI
(``EI-tri'') outperforms its multi-start numerical analog (``EI'') despite a
factor of five fewer evaluations of the EI criterion (about 9000 compared to
1700).  A more aggressive, gradient-based search misutilizes computational
resources. Finding a precise solution -- ultimately furnishing a maximal local
value in an inferior domain of attraction --  may come at the expense of
finding an accurate approximation near a global solution. The same is true,
but to a lesser extent, when comparing TS variations (5\% reduction).

See Appendix \ref{app:hart6} for nearly identical results with the 
higher dimensional Hartmann 6 function.

\subsection{Assemble to order (ATO)}
\label{sec:ato}

The assemble-to-order (ATO) simulator \citep{Hong:2006} deploys a queuing
system to explore profit for a manufacturer receiving orders for a variety of
widgets, each with different demand for component parts. Target inventory
levels for eight parts comprise the controllable inputs ($d=8$).  The output
is profit, a scalar distilling inventory and order fulfillment costs under
random orders for parts, fulfillment and inventory replenishment times.  ATO
is coded in {\sf Matlab} and exhibits input-dependent noise. Although it is
reasonably fast (seconds per evaluation), relatively large $n$ is required to
separate signal from noise ($n_0 = 80$ and $n_{\mathrm{end}} = 300$).  Cubic
computational bottlnecks for GP inference are in play here, despite using a
thrifty heteroskedastic GP \citep{binois2018practical}. More details,
including software, are provided in Appendix \ref{app:implement}. Since our
{\sf Matlab} licenses prevented us from running parallel instances, we had to
limit our experiment somewhat to obtain results in a reasonable amount of
time. Therefore we only entertained three acquisition alternatives: random
multi-start numerically optimized EI, tricands with the default
$N_{\mathrm{sub}} = 100d = 800$, and a hybrid using the best tricands
candidate to initialize a local numerically optimized EI (``EI-hyb'').

The bottom panels of Figure \ref{fig:gphart} track BOV for these three
alternatives in a view similar to the panels above.  Note that the original
ATO problem involves maximization, so we have negated the output and also
scaled it so that its units were more similar to our other benchmark problems.
Since the output is random, the BOVs are estimates.  We used the surrogate fit
at $n_{\mathrm{end}}$ to assign in-sample predictions via
$\mu_{n_{\mathrm{end}}}(x_i)$.  Otherwise, the story is similar to our earlier
synthetic results.  Tricands' progress is initially slower than numerically
optimized EI, but eventually gives lower BOV values.  The difference at
$n_{\mathrm{end}}$ is slight, but statistically significant.  A paired Wilcox
test with alternative hypothesis that BOV for ``EI-tri'' is below ``EI'' has a
$p$-value of 0.0037.  The hybrid is even better, beating pure tricands and
pure numerical optimization 85 and 88 out of 100 times, respectively.

\section{Sampling-based surrogates}
\label{sec:nonstat}

Tricands are most valuable when the inner-optimization problem cannot
be solved by library-based numerical methods, even locally. This
happens when the surrogate predictive surface is discontinuous and/or when
inference requires MCMC. 
% It is technically possible to optimize an objective
% based on approximate ergodic averages -- derivatives of sums are sums of
% derivatives, after all. One approach, leveraging common random numbers, is
% known sample average approximation \citep[SAA;][]{verweij2003sample}.
% However, implementation in practice is cumbersome and is not supported by any
% of the MCMC-based surrogate modeling libraries that we are aware of.
% Candidate-based acquisition is the {\em modus operandi} in this context.
%
% We have two specific examples in mind, both of which make for popular
% surrogate modeling choices when the 
Our examples here involve response surfaces exhibiting nonstationarity,
meaning that the input--output dynamics evolve over the input space. 
This demands a more elaborate surrogate.  We consider two.
% When
% those dynamics are not well-modeled by purely relative-distance-based
% structures, which are the default for GPs, one needs a more
% elaborate model. 
A treed Gaussian process \citep[TGP;][]{gramacy2008bayesian} uses axis-aligned
partitioning with GPs.  Such divide-and-conquer excels when dynamics change
abruptly across individual inputs creating distinct regimes.  MCMC can average
over the location of probable partition boundaries.  We use the {\sf R}
package {\tt tgp} on CRAN \citep{gramacy2007tgp}, following Section 4 of
\citet{gramacy2010categorical} for candidate-based EI acquisition.

Deep Gaussian processes \citep[DGPs;][]{damianou2013deep} warp inputs to
accommodate a more subtle evolution of dynamics in the input space compared to
the abrupt regime changes of TGP.  Although variational inference is popular
for DGPs \citep{salimbeni2017doubly}, \citet{sauer2020active}~argue that in
active learning contexts, such as BO, full posterior integration leads to
better uncertainty quantification, and thus  better acquisitions.  Here we use
the {\sf R} package {\tt deepgp} on CRAN \citep{deepgp} which supports EI
evaluation on candidates.

Unfortunately, neither {\tt tgp} nor {\tt deepgp} facilitate TS
acquisition for BO. 
% Taking a single draw from the posterior predictive
% distribution at a pre-defined set of candidates, but at the same time
% averaging over the posterior distribution of all unknowns, would require the
% use of common random numbers (like SAA for EI above).  
%
% because they don't allow the user any control over the random deviates
% deployed in sampling.  
% We therefore do not include TS in our comparisons here. 
However, the {\tt tgp} package furnishes a maximum {\em a posteriori} sample
which can be used as the basis of a local search \citep[][Section
4]{gramacy2010categorical}. 
Here we show that tricands provides better candidates for both pure
candidate EI search and this hybrid scheme.

\subsection{Abrupt changes}
\label{sec:abrupt}

The Gramacy \& Lee (G\&L) function benefits from a model supporting hard
breaks even though its dynamics evolve smoothly.
%\footnote{\url{http://www.sfu.ca/~ssurjano/grlee08.html}} 
The 2d input domain (coded to $[0,1]^2$) is mostly flat except in one quadrant
where a local maximum is twinned with a (global) minimum.
% , disguising it from the rest
% of the input space in some perspectives. 
The domain of attraction of that
global minimum covers only about 10\% of the input space. It is easily missed
by random initial designs and candidate sets.
% , which makes this a challenging
% BO benchmark.  
TGP is able to isolate the interesting quadrant with just two
axis aligned partitions, thus recognizing that sampling effort should be
concentrated there.

\begin{figure*}[ht!]
\centering
\includegraphics[scale=0.44,trim=5 10 25 10,clip=TRUE]{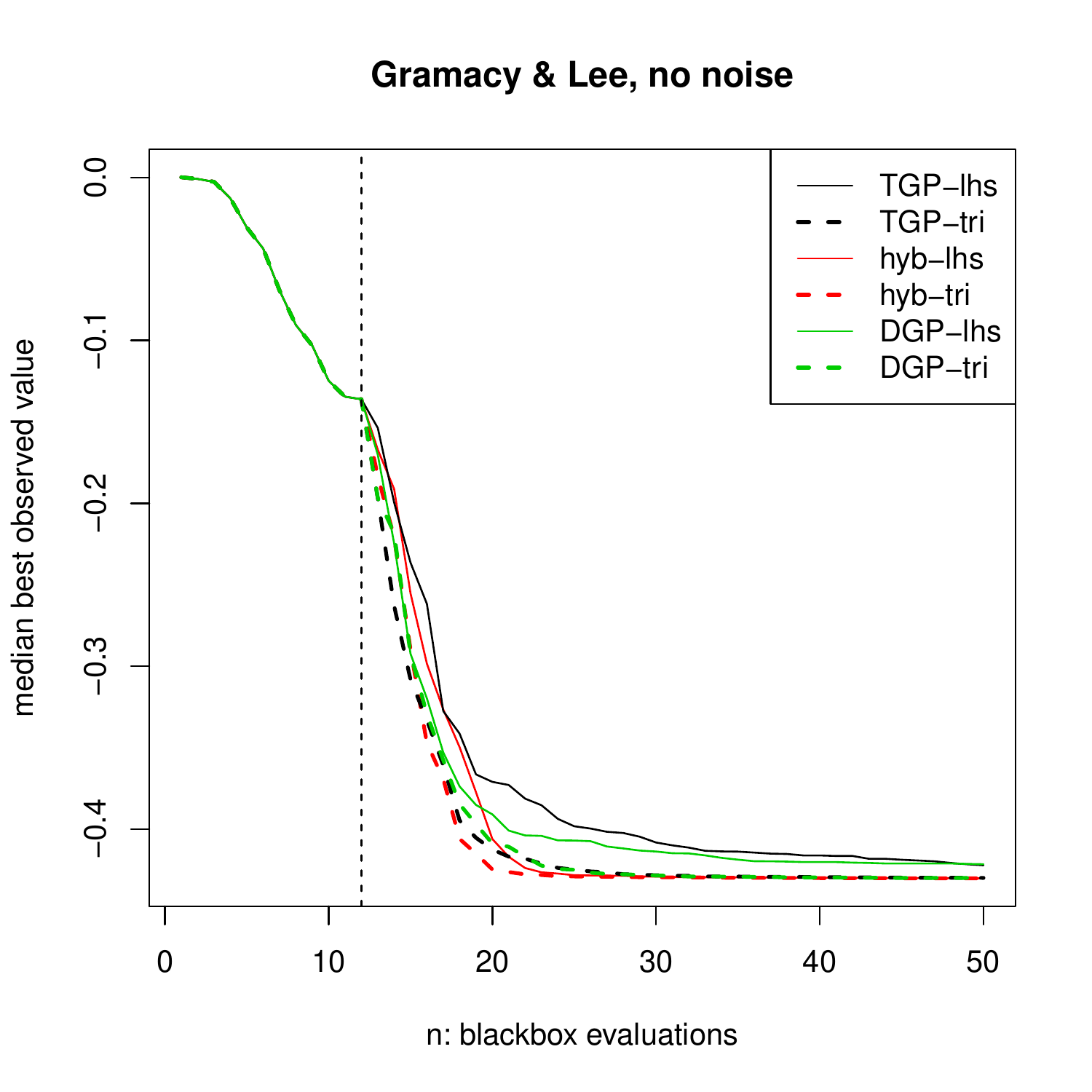}
\includegraphics[scale=0.44,trim=5 10 25 10,clip=TRUE]{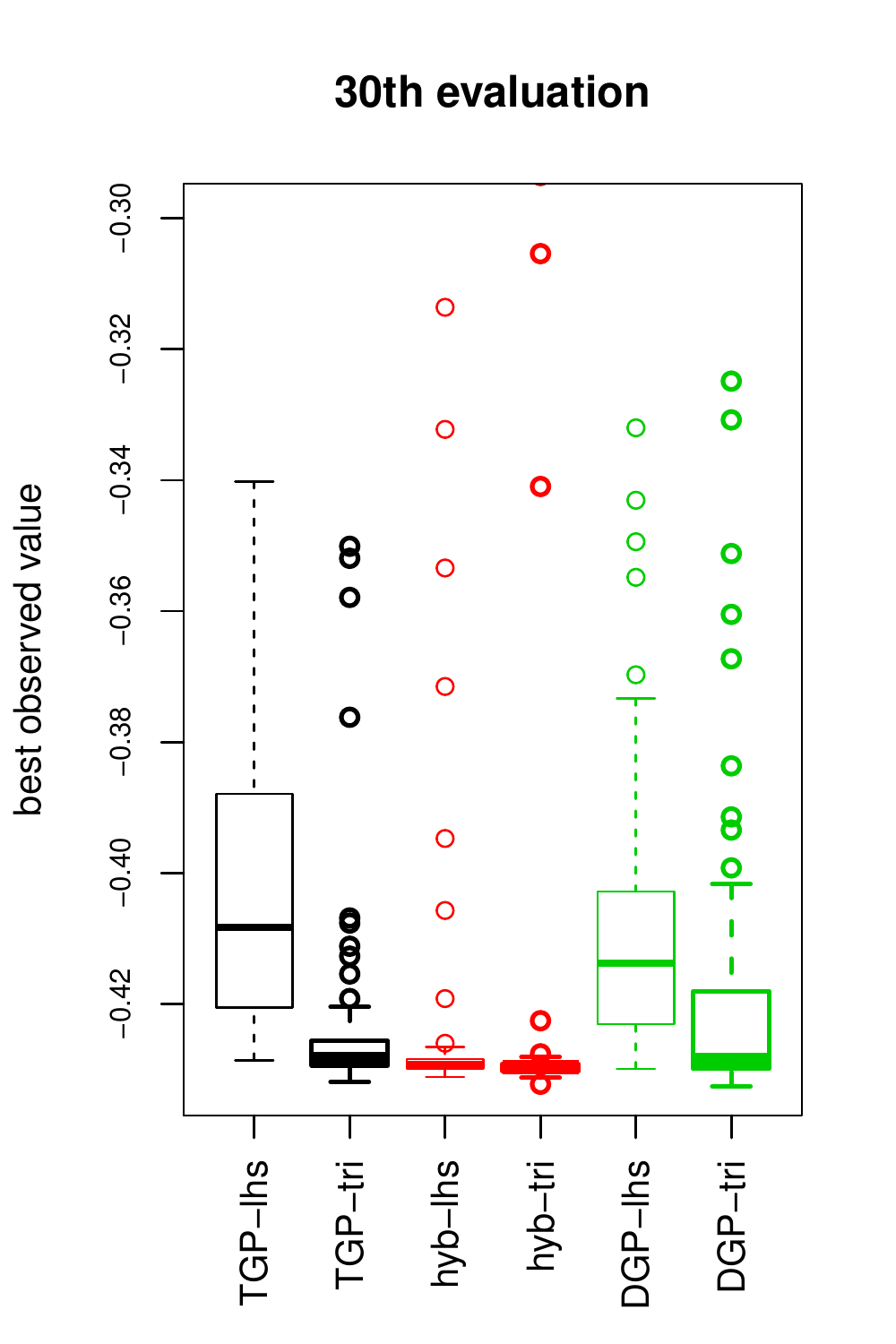}  
\includegraphics[scale=0.44,trim=50 10 20 10,clip=TRUE]{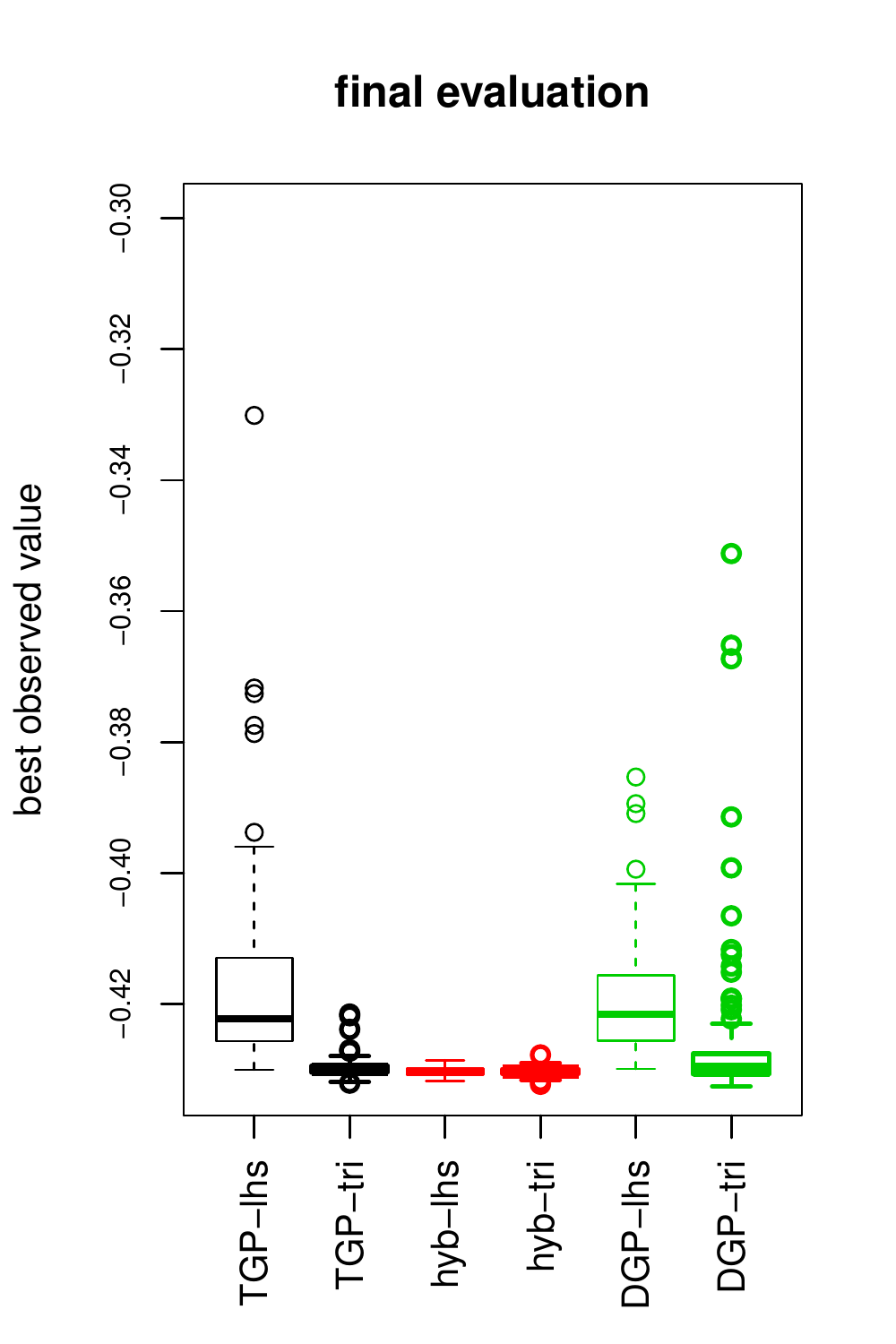} 
\includegraphics[scale=0.44,trim=5 10 25 10,clip=TRUE]{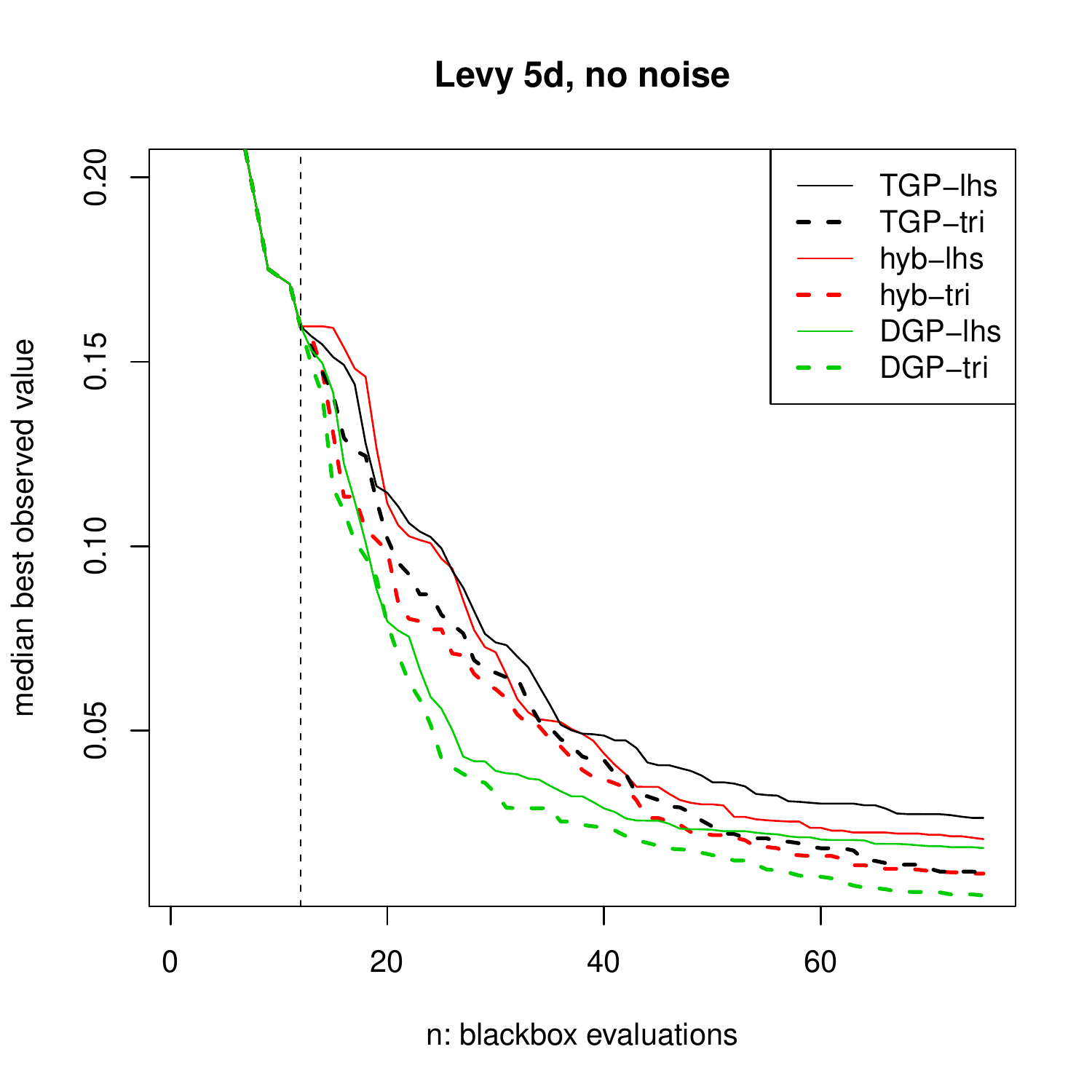}
\includegraphics[scale=0.44,trim=5 10 25 10,clip=TRUE]{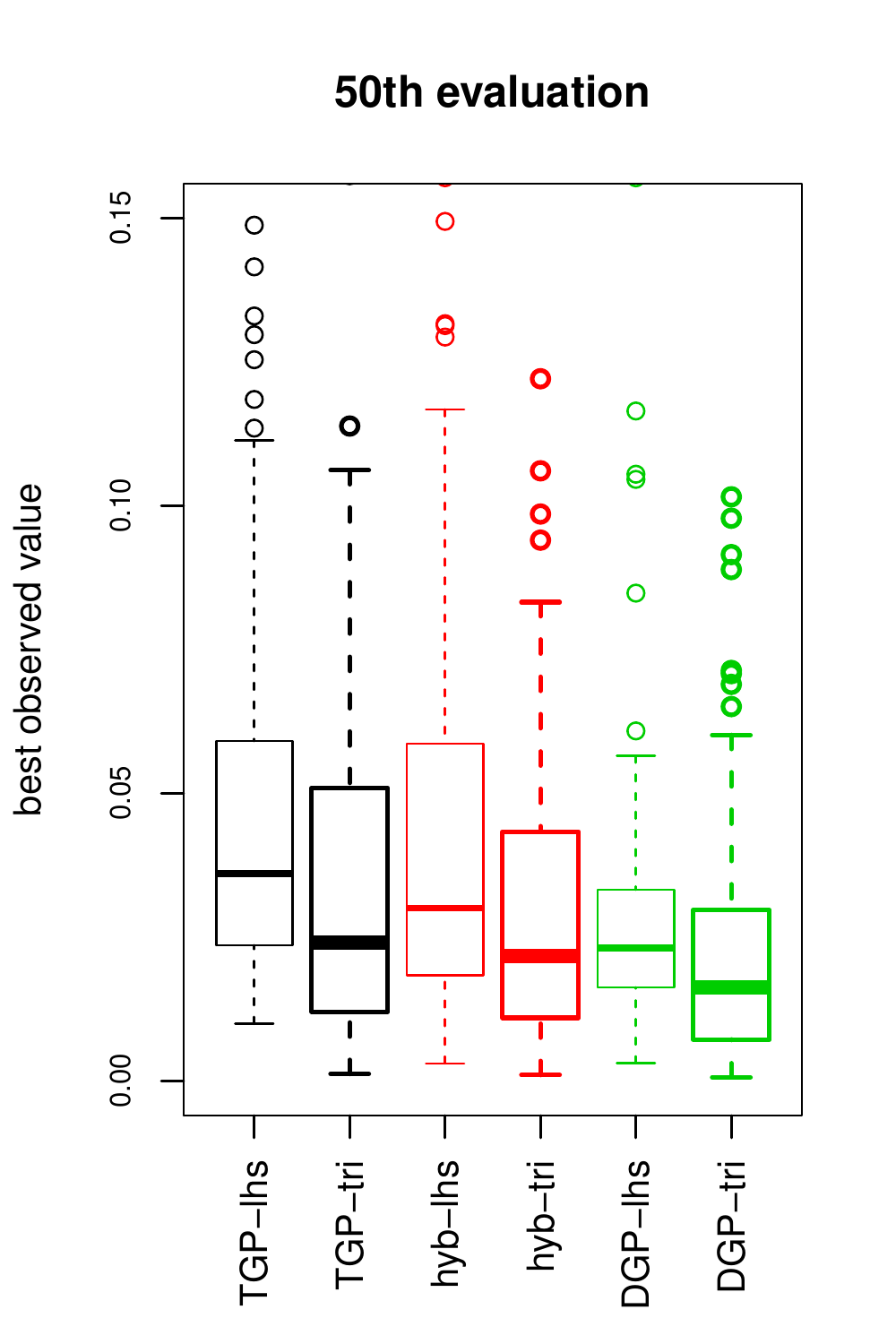} 
\includegraphics[scale=0.44,trim=50 10 20 10,clip=TRUE]{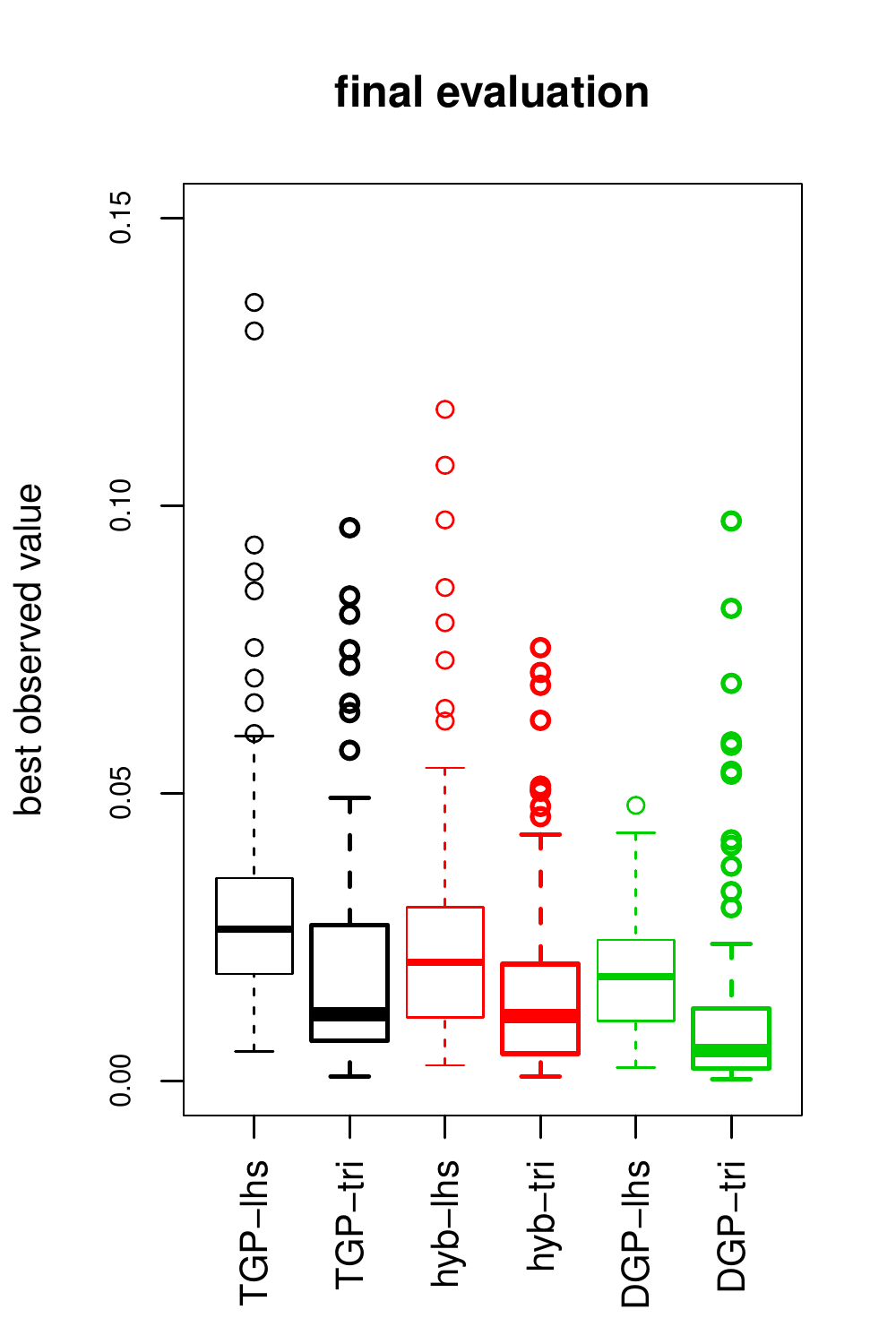} 
\vspace{-0.25cm}
\caption{Gramacy \& Lee (top) and 5d Levy (bottom) with MCMC-based comparators.
\label{fig:gllevy} 
}
\end{figure*}

The top row of Figure \ref{fig:gllevy} shows results in the same three views
as earlier. The number of candidates is limited to twenty, otherwise the setup
is unchanged. First ignore the red lines and boxplots (those labeled ``hyb-'')
and focus on the black (TGP) and green (DGP).  Notice that in both cases the
tricands-based comparators, dashed and/or bolded, outperform their LHS-based
analogues. In terms of medians over $n$ (top-left panel), that dominance is
uniform. In terms of boxplots, the disparity is stark with the exception of a
few outlying DGP-based BOV values with tricands.  TGP seems to edge out DGP
(with tricands) which we attribute to the abrupt change between the interesting
quadrant and the rest of the input space.  Now focus on the red ``hyb-''
results. These are the ones where TGP candidate-based search is finished with
a gradient-based EI search on the maximum {\em a posteriori} model.
Again, tricands wins. To supplement the visuals, we report that the tricands
hybrid BOV value was lowest in 97/100 reps (top-right).%  This is a clear win for tricands.

A higher dimensional example using the Michaelwicz function is
provided in Appendix \ref{app:mich}.

\subsection{Smooth changes}

The Levy function is defined in arbitrary
dimension. %\footnote{\url{http://www.sfu.ca/~ssurjano/levy.html}} 
It is ruffled
with many peaks and valleys.  Here we consider $d=5$ for variety; see the
bottom row of Figure \ref{fig:gllevy}. Commensurate with our other examples,
we consider two hundred candidates and $n_{\mathrm{end}}=75$.

% \begin{figure*}[ht!]
% \centering
% \includegraphics[scale=0.52,trim=5 10 25 10,clip=TRUE]{levy_prog}
% \includegraphics[scale=0.52,trim=5 10 25 10,clip=TRUE]{levy_50} 
% \includegraphics[scale=0.52,trim=50 10 20 10,clip=TRUE]{levy_75} 
% \vspace{-0.25cm}
% \caption{Levy function in 5d with MCMC-based comparators.
% \label{fig:levy5} 
% }
% \end{figure*}

Both candidate DGP variations outperform their TGP counterparts.
Slowly/smoothly varying dynamics favor warping inputs as opposed to hard
partitioning.  This is evident in both median and full distribution (boxplot)
views.  In the median view (bottom-left), notice that the dashed lines are
uniformly under the solid ones of the same color.  Tricands are providing
better median performance over all $n$. Although the preeminence of tricands
is apparent visually, we report that it provided better BOV in 81/100 MC
restarts among the best (DGP) comparators.  When tricands are involved, the
hybrid candidate/numerical option is not discernibly better than the pure
candidate alternative.% Its
% thrifty geometric criterion is an adequate substitute for, if not substantially
% superior to, aggressive acquisition criterion optimization.

\section{Discussion}
\label{sec:discuss}

We offer a novel take on space-filling candidates for acquisition in BO.  The
idea is to fill the spaces in-between previous acquisitions.  This is
motivated by an analogy to bisection search, but also by the nature of GP
predictive surfaces which often serve as surrogates in BO. GP surrogates have
organically inflated uncertainty between training data sites, which makes
those spots attractive for BO acquisition.  This notion is extended to the
space between the convex hull of existing training data and the boundary
$\mathcal{B}$ of the study region. In an array of benchmark exercises we have
demonstrated that these tricands lead to superior performance in BO compared
to both higher-powered gradient-based acquisition schemes and simpler
space-filling candidates.  Tricands' main attraction is that they mimic the
behavior of higher-powered searches with the implementation simplicity
of candidates.  That simplicity means that tricands may be deployed
where the higher-powered alternatives cannot, such as when the surrogate is
not continuous or requires MCMC.

Additional discussion on high input dimension and other extensions is
provided in Appendix \ref{app:discuss}.

\begin{ack}
We are grateful for funding from DOE LAB 17-1697 via
subaward from Argonne National Laboratory for SciDAC/DOE Office of Science
ASCR and High Energy Physics.
\end{ack}

%\newpage  % UNBLINDED
{
\small
\bibliographystyle{plainnat}
\bibliography{tricands}

\begin{thebibliography}{65}
\providecommand{\natexlab}[1]{#1}
\providecommand{\url}[1]{\texttt{#1}}
\expandafter\ifx\csname urlstyle\endcsname\relax
  \providecommand{\doi}[1]{doi: #1}\else
  \providecommand{\doi}{doi: \begingroup \urlstyle{rm}\Url}\fi

\bibitem[Azzimonti et~al.(2020)Azzimonti, Ginsbourger, Chevalier, Bect, and
  Richet]{azzimonti2020adaptive}
Dario Azzimonti, David Ginsbourger, Cl{\'e}ment Chevalier, Julien Bect, and
  Yann Richet.
\newblock Adaptive design of experiments for conservative estimation of
  excursion sets.
\newblock \emph{Technometrics}, pages 1--14, 2020.

\bibitem[Baker et~al.(2020)Baker, Barbillon, Fadikar, Gramacy, Herbei, Higdon,
  Huang, Johnson, Ma, Mondal, et~al.]{baker2020analyzing}
Evan Baker, Pierre Barbillon, Arindam Fadikar, Robert~B Gramacy, Radu Herbei,
  David Higdon, Jiangeng Huang, Leah~R Johnson, Pulong Ma, Anirban Mondal,
  et~al.
\newblock Analyzing stochastic computer models: A review with opportunities.
\newblock \emph{arXiv preprint arXiv:2002.01321}, 2020.

\bibitem[Barber et~al.(1996)Barber, Dobkin, and Huhdanpaa]{quickhull}
C.~Bradford Barber, David~P. Dobkin, and Hannu Huhdanpaa.
\newblock The quickhull algorithm for convex hulls.
\newblock \emph{ACM Trans. Math. Softw.}, 22\penalty0 (4):\penalty0 469–483,
  December 1996.
\newblock ISSN 0098-3500.
\newblock \doi{10.1145/235815.235821}.
\newblock URL \url{https://doi.org/10.1145/235815.235821}.

\bibitem[Bates and Pronzato(2001)]{bates2001tri}
R.~Bates and L.~Pronzato.
\newblock Emulator-based global optimisation using lattices and delaunay
  tesselation.
\newblock In \emph{Sensitivity Analysis of Model Output}, pages 189--192,
  Madrid (Espagne), 2001.

\bibitem[Bect et~al.(2012)Bect, Ginsbourger, Li, Picheny, and
  Vazquez]{bect2012sequential}
Julien Bect, David Ginsbourger, Ling Li, Victor Picheny, and Emmanuel Vazquez.
\newblock Sequential design of computer experiments for the estimation of a
  probability of failure.
\newblock \emph{Statistics and Computing}, 22\penalty0 (3):\penalty0 773--793,
  2012.

\bibitem[Bengtsson(2018)]{R.matlab}
Henrik Bengtsson.
\newblock \emph{R.matlab: Read and Write {MAT} Files and Call {MATLAB} from
  Within {R}}, 2018.
\newblock URL \url{https://CRAN.R-project.org/package=R.matlab}.
\newblock R package version 3.6.2.

\bibitem[Bergstra et~al.(2011)Bergstra, Bardenet, Bengio, and
  K{\'e}gl]{bergstra2011algorithms}
James Bergstra, R{\'e}mi Bardenet, Yoshua Bengio, and Bal{\'a}zs K{\'e}gl.
\newblock Algorithms for hyper-parameter optimization.
\newblock \emph{Advances in neural information processing systems}, 24, 2011.

\bibitem[Binois et~al.(2018)Binois, Gramacy, and
  Ludkovski]{binois2018practical}
M~Binois, RB~Gramacy, and M~Ludkovski.
\newblock Practical heteroscedastic {G}aussian process modeling for large
  simulation experiments.
\newblock \emph{Journal of Computational and Graphical Statistics}, 27\penalty0
  (4):\penalty0 808--821, 2018.
\newblock \doi{10.1080/10618600.2018.1458625}.
\newblock URL \url{https://doi.org/10.1080/10618600.2018.1458625}.

\bibitem[Binois et~al.(2019)Binois, Huang, Gramacy, and
  Ludkovski]{binois2018replication}
M~Binois, J~Huang, RB~Gramacy, and M~Ludkovski.
\newblock Replication or exploration? {S}equential design for stochastic
  simulation experiments.
\newblock \emph{Technometrics}, 27\penalty0 (4):\penalty0 808--821, 2019.
\newblock \doi{10.1080/00401706.2018.1469433}.
\newblock URL \url{https://doi.org/10.1080/00401706.2018.1469433}.

\bibitem[Binois and Wycoff(2021)]{binois2021survey}
Mickael Binois and Nathan Wycoff.
\newblock A survey on high-dimensional {G}aussian process modeling with
  application to bayesian optimization, 2021.

\bibitem[Binois and Gramacy(2021)]{JSSv098i13}
Mickaël Binois and Robert~B. Gramacy.
\newblock hetgp: Heteroskedastic {G}aussian process modeling and sequential
  design in {\sf r}.
\newblock \emph{Journal of Statistical Software}, 98\penalty0 (13):\penalty0
  1–44, 2021.
\newblock \doi{10.18637/jss.v098.i13}.
\newblock URL
  \url{https://www.jstatsoft.org/index.php/jss/article/view/v098i13}.

\bibitem[Breiman(2001)]{breiman2001random}
Leo Breiman.
\newblock Random forests.
\newblock \emph{Machine learning}, 45\penalty0 (1):\penalty0 5--32, 2001.

\bibitem[Bull(2011)]{bull2011convergence}
AD~Bull.
\newblock Convergence rates of efficient global optimization algorithms.
\newblock \emph{Journal of Machine Learning Research}, 12\penalty0
  (Oct):\penalty0 2879--2904, 2011.

\bibitem[Burden and Faires(1985)]{burden1985numerical}
R.L. Burden and J.D. Faires.
\newblock \emph{Numerical Analysis}.
\newblock PWS Publishers, 1985.

\bibitem[Byrd et~al.(2003)Byrd, Lu, Nocedal, and Zhu]{BFGS}
Richardh Byrd, Peihuang Lu, Jorge Nocedal, and Ciyou Zhu.
\newblock A limited memory algorithm for bound constrained optimization.
\newblock \emph{SIAM Journal on Scientific Computing}, 16, 02 2003.
\newblock \doi{10.1137/0916069}.

\bibitem[Carnell(2018)]{R-lhs}
Rob Carnell.
\newblock \emph{{\tt lhs}: {L}atin Hypercube Samples}, 2018.
\newblock URL \url{https://CRAN.R-project.org/package=lhs}.
\newblock R package version 0.16.

\bibitem[Chevalier and Ginsbourger(2013)]{chevalier2013fast}
Cl{\'e}ment Chevalier and David Ginsbourger.
\newblock Fast computation of the multi-points expected improvement with
  applications in batch selection.
\newblock In \emph{International Conference on Learning and Intelligent
  Optimization}, pages 59--69. Springer, 2013.

\bibitem[Chevalier et~al.(2014)Chevalier, Bect, Ginsbourger, Vazquez, Picheny,
  and Richet]{chevalier2014fast}
Cl{\'e}ment Chevalier, Julien Bect, David Ginsbourger, Emmanuel Vazquez, Victor
  Picheny, and Yann Richet.
\newblock Fast parallel kriging-based stepwise uncertainty reduction with
  application to the identification of an excursion set.
\newblock \emph{Technometrics}, 56\penalty0 (4):\penalty0 455--465, 2014.

\bibitem[Cole et~al.(2021)Cole, Gramacy, Warner, Bomarito, Leser, and
  Leser]{cole2021entropy}
D~Austin Cole, Robert~B Gramacy, James~E Warner, Geoffrey~F Bomarito, Patrick~E
  Leser, and William~P Leser.
\newblock Entropy-based adaptive design for contour finding and estimating
  reliability.
\newblock \emph{arXiv preprint arXiv:2105.11357}, 2021.

\bibitem[Damianou and Lawrence(2013)]{damianou2013deep}
Andreas Damianou and Neil~D Lawrence.
\newblock Deep {G}aussian processes.
\newblock In \emph{Artificial intelligence and statistics}, pages 207--215.
  PMLR, 2013.

\bibitem[Daulton et~al.(2021)Daulton, Eriksson, Balandat, and
  Bakshy]{daulton2021multiobjective}
Samuel Daulton, David Eriksson, Maximilian Balandat, and Eytan Bakshy.
\newblock Multi-objective {B}ayesian optimization over high-dimensional search
  spaces, 2021.

\bibitem[Eriksson et~al.(2019)Eriksson, Pearce, Gardner, Turner, and
  Poloczek]{eriksson2019scalable}
David Eriksson, Michael Pearce, Jacob Gardner, Ryan~D Turner, and Matthias
  Poloczek.
\newblock Scalable global optimization via local {B}ayesian optimization.
\newblock In \emph{Advances in Neural Information Processing Systems}, pages
  5497--5508, 2019.

\bibitem[Feurer et~al.(2018)Feurer, Letham, Hutter, and
  Bakshy]{feurer2018practical}
Matthias Feurer, Benjamin Letham, Frank Hutter, and Eytan Bakshy.
\newblock Practical transfer learning for bayesian optimization.
\newblock \emph{arXiv preprint arXiv:1802.02219}, 2018.

\bibitem[Frazier et~al.(2008)Frazier, Powell, and
  Dayanik]{frazier2008knowledge}
Peter~I Frazier, Warren~B Powell, and Savas Dayanik.
\newblock A knowledge-gradient policy for sequential information collection.
\newblock \emph{SIAM Journal on Control and Optimization}, 47\penalty0
  (5):\penalty0 2410--2439, 2008.

\bibitem[Garnett(2022)]{garnett_bayesoptbook_2022}
Roman Garnett.
\newblock \emph{{Bayesian Optimization}}.
\newblock Cambridge University Press, 2022.
\newblock in preparation.

\bibitem[Ginsbourger et~al.(2007)Ginsbourger, Le~Riche, and
  Carraro]{ginsbourger2007multi}
David Ginsbourger, Rodolphe Le~Riche, and Laurent Carraro.
\newblock A multi-points criterion for deterministic parallel global
  optimization based on kriging.
\newblock In \emph{NCP07}, 2007.

\bibitem[Gonzalez et~al.(2016)Gonzalez, Osborne, and
  Lawrence]{gonzalez2016glasses}
J~Gonzalez, M~Osborne, and N~Lawrence.
\newblock {\tt GLASSES}: {R}elieving the myopia of {B}ayesian optimisation.
\newblock In \emph{Proceedings of the 19th International Conference on
  Artificial Intelligence and Statistics}, pages 790--799, 2016.

\bibitem[Gonz{\'a}lez et~al.(2016)Gonz{\'a}lez, Dai, Hennig, and
  Lawrence]{gonzalez2016batch}
Javier Gonz{\'a}lez, Zhenwen Dai, Philipp Hennig, and Neil Lawrence.
\newblock Batch bayesian optimization via local penalization.
\newblock In \emph{Artificial intelligence and statistics}, pages 648--657.
  PMLR, 2016.

\bibitem[Gramacy(2007)]{gramacy2007tgp}
RB~Gramacy.
\newblock {\tt tgp}: an {R} package for {B}ayesian nonstationary,
  semiparametric nonlinear regression and design by treed {G}aussian process
  models.
\newblock \emph{Journal of Statistical Software}, 19\penalty0 (9):\penalty0 6,
  2007.

\bibitem[Gramacy(2016)]{gramacy2016lagp}
RB~Gramacy.
\newblock {\tt laGP}: large-scale spatial modeling via local approximate
  {G}aussian processes in {R}.
\newblock \emph{Journal of Statistical Software}, 72\penalty0 (1):\penalty0
  1--46, 2016.

\bibitem[Gramacy and Lee(2011)]{gramacy2011optimization}
RB~Gramacy and Herbert~KH Lee.
\newblock Optimization under unknown constraints.
\newblock In \emph{Bayesian Statistics}, volume~9. Oxford University Press,
  2011.

\bibitem[Gramacy and Lee(2008)]{gramacy2008bayesian}
RB~Gramacy and HKH Lee.
\newblock Bayesian treed {G}aussian process models with an application to
  computer modeling.
\newblock \emph{Journal of the American Statistical Association}, 103\penalty0
  (483):\penalty0 1119--1130, 2008.

\bibitem[Gramacy and Sun(2018)]{R-laGP}
RB~Gramacy and F~Sun.
\newblock \emph{{\tt laGP}: {L}ocal Approximate {G}aussian Process Regression},
  2018.
\newblock URL \url{http://bobby.gramacy.com/r_packages/laGP}.
\newblock R package version 1.5-3.

\bibitem[Gramacy and Taddy(2010)]{gramacy2010categorical}
RB~Gramacy and MA~Taddy.
\newblock Categorical inputs, sensitivity analysis, optimization and importance
  tempering with {\tt tgp} version 2, an {R} package for treed {G}aussian
  process models.
\newblock \emph{Journal of Statistical Software}, 33\penalty0 (6):\penalty0
  1--48, 2010.

\bibitem[Gramacy(2020)]{gramacy2020surrogates}
Robert~B. Gramacy.
\newblock \emph{Surrogates: {G}aussian Process Modeling, Design and
  Optimization for the Applied Sciences}.
\newblock Chapman Hall/CRC, Boca Raton, Florida, 2020.
\newblock \url{http://bobby.gramacy.com/surrogates/}.

\bibitem[Habel et~al.(2019)Habel, Grasman, Gramacy, Mozharovskyi, and
  Sterratt]{geometry}
Kai Habel, Raoul Grasman, Robert~B. Gramacy, Pavlo Mozharovskyi, and David~C.
  Sterratt.
\newblock \emph{{\tt geometry}: Mesh Generation and Surface Tessellation},
  2019.
\newblock URL \url{https://CRAN.R-project.org/package=geometry}.
\newblock R package version 0.4.5.

\bibitem[Hong and Nelson(2006)]{Hong:2006}
L.J. Hong and B.L. Nelson.
\newblock Discrete optimization via simulation using compass.
\newblock \emph{Operations Research}, 54\penalty0 (1):\penalty0 115--129, 2006.

\bibitem[Jones et~al.(1998)Jones, Schonlau, and Welch]{jones1998efficient}
Donald~R Jones, Matthias Schonlau, and William~J Welch.
\newblock Efficient global optimization of expensive black-box functions.
\newblock \emph{Journal of Global optimization}, 13\penalty0 (4):\penalty0
  455--492, 1998.

\bibitem[Lam et~al.(2016)Lam, Willcox, and Wolpert]{lam2016bayesian}
Remi Lam, Karen Willcox, and David~H Wolpert.
\newblock Bayesian optimization with a finite budget: An approximate dynamic
  programming approach.
\newblock \emph{Advances in Neural Information Processing Systems},
  29:\penalty0 883--891, 2016.

\bibitem[Leatherman et~al.(2017)Leatherman, Santner, and
  Dean]{leatherman2017computer}
ER~Leatherman, TJ~Santner, and AM~Dean.
\newblock Computer experiment designs for accurate prediction.
\newblock \emph{Statistics and Computing}, pages 1--13, 2017.

\bibitem[Lee and Schachter(1980)]{lee1980two}
Der-Tsai Lee and Bruce~J Schachter.
\newblock Two algorithms for constructing a delaunay triangulation.
\newblock \emph{International Journal of Computer \& Information Sciences},
  9\penalty0 (3):\penalty0 219--242, 1980.

\bibitem[Letham and Bakshy(2019)]{letham2019bayesian}
Benjamin Letham and Eytan Bakshy.
\newblock Bayesian optimization for policy search via online-offline
  experimentation.
\newblock \emph{J. Mach. Learn. Res.}, 20:\penalty0 145--1, 2019.

\bibitem[Marques et~al.(2018)Marques, Lam, and Willcox]{marques2018contour}
Alexandre Marques, Remi Lam, and Karen Willcox.
\newblock Contour location via entropy reduction leveraging multiple
  information sources.
\newblock In \emph{Advances in Neural Information Processing Systems}, pages
  5217--5227, 2018.

\bibitem[Mckay et~al.(1979)Mckay, Beckman, and Conover]{Mckay:1979}
D~Mckay, Richard Beckman, and William Conover.
\newblock A comparison of three methods for selecting vales of input variables
  in the analysis of output from a computer code.
\newblock \emph{Technometrics}, 21:\penalty0 239--245, 05 1979.

\bibitem[Mo{\v{c}}kus(1975)]{movckus1975bayesian}
Jonas Mo{\v{c}}kus.
\newblock On bayesian methods for seeking the extremum.
\newblock In \emph{Optimization techniques IFIP technical conference}, pages
  400--404. Springer, 1975.

\bibitem[Nelder and Mead(1965)]{nelder1965simplex}
JA~Nelder and R~Mead.
\newblock A simplex method for function minimization.
\newblock \emph{The Computer Journal}, 7\penalty0 (4):\penalty0 308--313, 1965.

\bibitem[Paszke et~al.(2017)Paszke, Gross, Chintala, Chanan, Yang, DeVito, Lin,
  Desmaison, Antiga, and Lerer]{paszke2017automatic}
Adam Paszke, Sam Gross, Soumith Chintala, Gregory Chanan, Edward Yang, Zachary
  DeVito, Zeming Lin, Alban Desmaison, Luca Antiga, and Adam Lerer.
\newblock Automatic differentiation in pytorch.
\newblock 2017.

\bibitem[Picheny et~al.(2012)Picheny, Wagner, and
  Ginsbourger]{picheny2012benchmark}
V~Picheny, T~Wagner, and D~Ginsbourger.
\newblock A benchmark of kriging-based infill criteria for noisy optimization.
\newblock \emph{Structural and Multidisciplinary Optimization}, pages 1--20,
  2012.

\bibitem[Pourmohamad and Lee(2021)]{pourmohamadbayesian}
Tony Pourmohamad and Herbert~KH Lee.
\newblock \emph{Bayesian Optimization with Application to Computer
  Experiments}.
\newblock Springer, New York, NY, 2021.

\bibitem[Ranjan et~al.(2008)Ranjan, Bingham, and
  Michailidis]{ranjan2008sequential}
Pritam Ranjan, Derek Bingham, and George Michailidis.
\newblock Sequential experiment design for contour estimation from complex
  computer codes.
\newblock \emph{Technometrics}, 50\penalty0 (4):\penalty0 527--541, 2008.

\bibitem[Salimbeni and Deisenroth(2017)]{salimbeni2017doubly}
Hugh Salimbeni and Marc Deisenroth.
\newblock Doubly stochastic variational inference for deep {G}aussian
  processes.
\newblock \emph{arXiv preprint arXiv:1705.08933}, 2017.

\bibitem[Sauer(2021)]{deepgp}
Annie Sauer.
\newblock \emph{{\tt deepgp}: Sequential Design for Deep {G}aussian Processes
  using MCMC}, 2021.
\newblock URL \url{https://CRAN.R-project.org/package=deepgp}.
\newblock R package version 0.3.0.

\bibitem[Sauer et~al.(2020)Sauer, Gramacy, and Higdon]{sauer2020active}
Annie Sauer, Robert~B Gramacy, and David Higdon.
\newblock Active learning for deep {G}aussian process surrogates.
\newblock \emph{arXiv preprint arXiv:2012.08015}, 2020.

\bibitem[Scott et~al.(2011)Scott, Frazier, and Powell]{scott2011correlated}
Warren Scott, Peter Frazier, and Warren Powell.
\newblock The correlated knowledge gradient for simulation optimization of
  continuous parameters using {G}aussian process regression.
\newblock \emph{SIAM Journal on Optimization}, 21\penalty0 (3):\penalty0
  996--1026, 2011.

\bibitem[Srinivas et~al.(2009)Srinivas, Krause, Kakade, and
  Seeger]{srinivas2009gaussian}
Niranjan Srinivas, Andreas Krause, Sham~M Kakade, and Matthias Seeger.
\newblock Gaussian process optimization in the bandit setting: No regret and
  experimental design.
\newblock \emph{arXiv preprint arXiv:0912.3995}, 2009.

\bibitem[Su and Drysdale(1997)]{su1997comparison}
Peter Su and Robert L~Scot Drysdale.
\newblock A comparison of sequential delaunay triangulation algorithms.
\newblock \emph{Computational Geometry}, 7\penalty0 (5-6):\penalty0 361--385,
  1997.

\bibitem[Surjanovic and Bingham(2013)]{surjanovic2013virtual}
S~Surjanovic and D~Bingham.
\newblock Virtual library of simulation experiments: test functions and
  datasets.
\newblock \url{http://www.sfu.ca/~ssurjano}, 2013.

\bibitem[Taddy et~al.(2009)Taddy, Lee, Gray, and Griffin]{taddy2009bayesian}
MA~Taddy, HKH Lee, GA~Gray, and JD~Griffin.
\newblock Bayesian guided pattern search for robust local optimization.
\newblock \emph{Technometrics}, 51\penalty0 (4):\penalty0 389--401, 2009.

\bibitem[Thompson(1933)]{thompson1933likelihood}
William~R Thompson.
\newblock On the likelihood that one unknown probability exceeds another in
  view of the evidence of two samples.
\newblock \emph{Biometrika}, 25\penalty0 (3/4):\penalty0 285--294, 1933.

\bibitem[Turner et~al.(2021)Turner, Eriksson, McCourt, Kiili, Laaksonen, Xu,
  and Guyon]{turner2021bayesian}
Ryan Turner, David Eriksson, Michael McCourt, Juha Kiili, Eero Laaksonen, Zhen
  Xu, and Isabelle Guyon.
\newblock Bayesian optimization is superior to random search for machine
  learning hyperparameter tuning: Analysis of the black-box optimization
  challenge 2020.
\newblock \emph{arXiv preprint arXiv:2104.10201}, 2021.

\bibitem[Virtanen et~al.(2020)Virtanen, Gommers, Oliphant, Haberland, Reddy,
  Cournapeau, Burovski, Peterson, Weckesser, Bright, {van der Walt}, Brett,
  Wilson, Millman, Mayorov, Nelson, Jones, Kern, Larson, Carey, Polat, Feng,
  Moore, {VanderPlas}, Laxalde, Perktold, Cimrman, Henriksen, Quintero, Harris,
  Archibald, Ribeiro, Pedregosa, {van Mulbregt}, and {SciPy 1.0
  Contributors}]{2020SciPy-NMeth}
Pauli Virtanen, Ralf Gommers, Travis~E. Oliphant, Matt Haberland, Tyler Reddy,
  David Cournapeau, Evgeni Burovski, Pearu Peterson, Warren Weckesser, Jonathan
  Bright, St{\'e}fan~J. {van der Walt}, Matthew Brett, Joshua Wilson, K.~Jarrod
  Millman, Nikolay Mayorov, Andrew R.~J. Nelson, Eric Jones, Robert Kern, Eric
  Larson, C~J Carey, {\.I}lhan Polat, Yu~Feng, Eric~W. Moore, Jake
  {VanderPlas}, Denis Laxalde, Josef Perktold, Robert Cimrman, Ian Henriksen,
  E.~A. Quintero, Charles~R. Harris, Anne~M. Archibald, Ant{\^o}nio~H. Ribeiro,
  Fabian Pedregosa, Paul {van Mulbregt}, and {SciPy 1.0 Contributors}.
\newblock {{SciPy} 1.0: Fundamental Algorithms for Scientific Computing in
  Python}.
\newblock \emph{Nature Methods}, 17:\penalty0 261--272, 2020.
\newblock \doi{10.1038/s41592-019-0686-2}.

\bibitem[Wang et~al.(2020)Wang, Fonseca, and Tian]{wang2020learning}
Linnan Wang, Rodrigo Fonseca, and Yuandong Tian.
\newblock Learning search space partition for black-box optimization using
  monte carlo tree search.
\newblock \emph{Advances in Neural Information Processing Systems},
  33:\penalty0 19511--19522, 2020.

\bibitem[Zhang et~al.(2020{\natexlab{a}})Zhang, Gramacy, Johnson, Rose, and
  Smith]{zhang2020batch}
Boya Zhang, Robert~B Gramacy, Leah Johnson, Kenneth~A Rose, and Eric Smith.
\newblock Batch-sequential design and heteroskedastic surrogate modeling for
  delta smelt conservation.
\newblock \emph{arXiv preprint arXiv:2010.06515}, 2020{\natexlab{a}}.

\bibitem[Zhang et~al.(2021)Zhang, Cole, and Gramacy]{zhang2021distance}
Boya Zhang, D~Austin Cole, and Robert~B Gramacy.
\newblock Distance-distributed design for {G}aussian process surrogates.
\newblock \emph{Technometrics}, 63\penalty0 (1):\penalty0 40--52, 2021.

\bibitem[Zhang et~al.(2020{\natexlab{b}})Zhang, Apley, and
  Chen]{zhang2020bayesian}
Yichi Zhang, Daniel~W Apley, and Wei Chen.
\newblock Bayesian optimization for materials design with mixed quantitative
  and qualitative variables.
\newblock \emph{Scientific reports}, 10\penalty0 (1):\penalty0 1--13,
  2020{\natexlab{b}}.

\end{thebibliography}
}

% UBLINDED

%%%%%%%%%%%%%%%%%%%%%%%%%%%%%%%%%%%%%%%%%%%%%%%%%%%%%%%%%%%%

\newpage 
\appendix

%\section{Appendix}
\section{Implementation and additional empirical results}
\label{app:results}

Here we summarize implementation details and experimental results
that were removed from the main body of the paper due to space constraints.
All of the empirical work in our paper is fully reproducible.  Code may be
found in our git repository: \url{http://bitbucket.org/gramacylab/tricands}.
%\url{http://bitbucket.org/blinded/tricands}.  UBLINDED

\subsection{Classical GP implementation details}
\label{app:implement}

The GP surrogate for the Goldstein--Price and Hartman 6 examples (Section
\ref{sec:classic} and Appendix \ref{app:hart6}, respectively) used the {\tt laGP}
package for {\sf R} \citep{gramacy2016lagp,R-laGP} on CRAN with
separable/automatic-relevance determination lengthscales in a Gaussian kernel
using gradient-based MLE.  We supplied our own EI function, following the
prototype provided in Chapter 7 of \citet{gramacy2020surrogates}, modified to
keep track of the number of evaluations.   Finite-differencing was used to 
approximate gradients.  For Latin hypercube samples we used
the {\tt lhs} \citep{R-lhs} package for {\sf R}.  ``Raw''
comparators L-BFGS-B and Nelder--Mead \citep{nelder1965simplex} local
optimizers were facilitated via the {\tt optim} function in {\sf R}.  

The heteroskedastic GP surrogate used for ATO (Section \ref{sec:ato}) was via
{\tt hetGP} \citep{JSSv098i13} for {\sf R}.  The built-in EI capability in
that package, which supports both numerical optimization (via analytic
gradients) and candidates, unfortunately could not easily be modified to
keep track of the number of evaluations.  Simulations, which require {\sf
Matlab}, were run in {\sf R} through the {\tt R.matlab} interface
\citep{R.matlab}. The input space for ATO is discrete, in $\{1,\dots,20\}^8$,
which is different from the real-valued inputs of our other examples.  The
only modification we made to accommodate this nuance was to ``snap''
acquisitions to that grid, implemented in coded units.  Sometimes this resulted
in replications in the design. No additional consideration was given to
replicates, despite findings that they are beneficial in similar contexts
\citep{binois2018replication}.

Details for the surrogates used in Section \ref{sec:nonstat} are provided
therein.

\subsection{Hartmann 6}
\label{app:hart6}

\begin{figure*}[ht!]
\centering
\includegraphics[scale=0.44,trim=5 10 25 10,clip=TRUE]{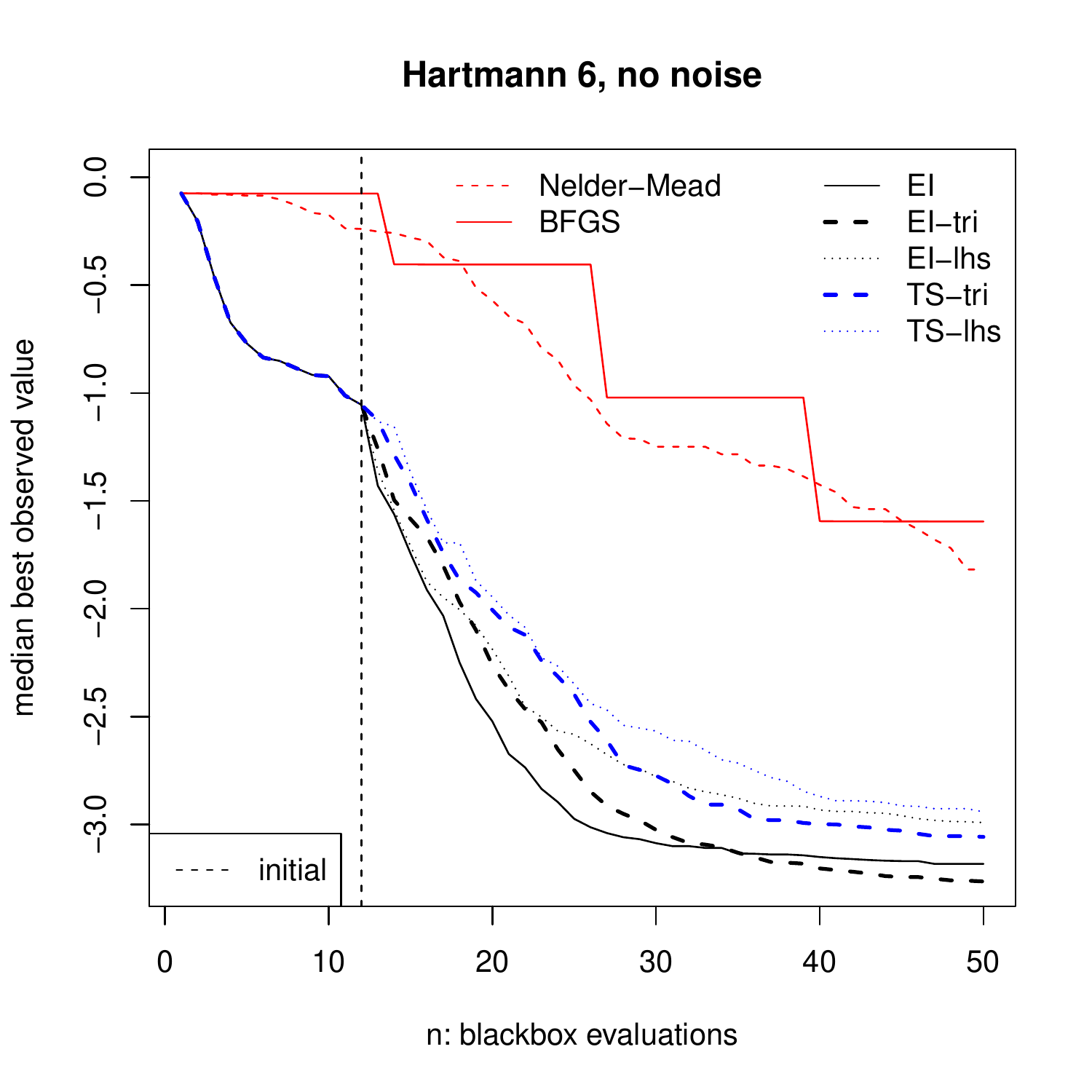}
\includegraphics[scale=0.44,trim=5 10 25 10,clip=TRUE]{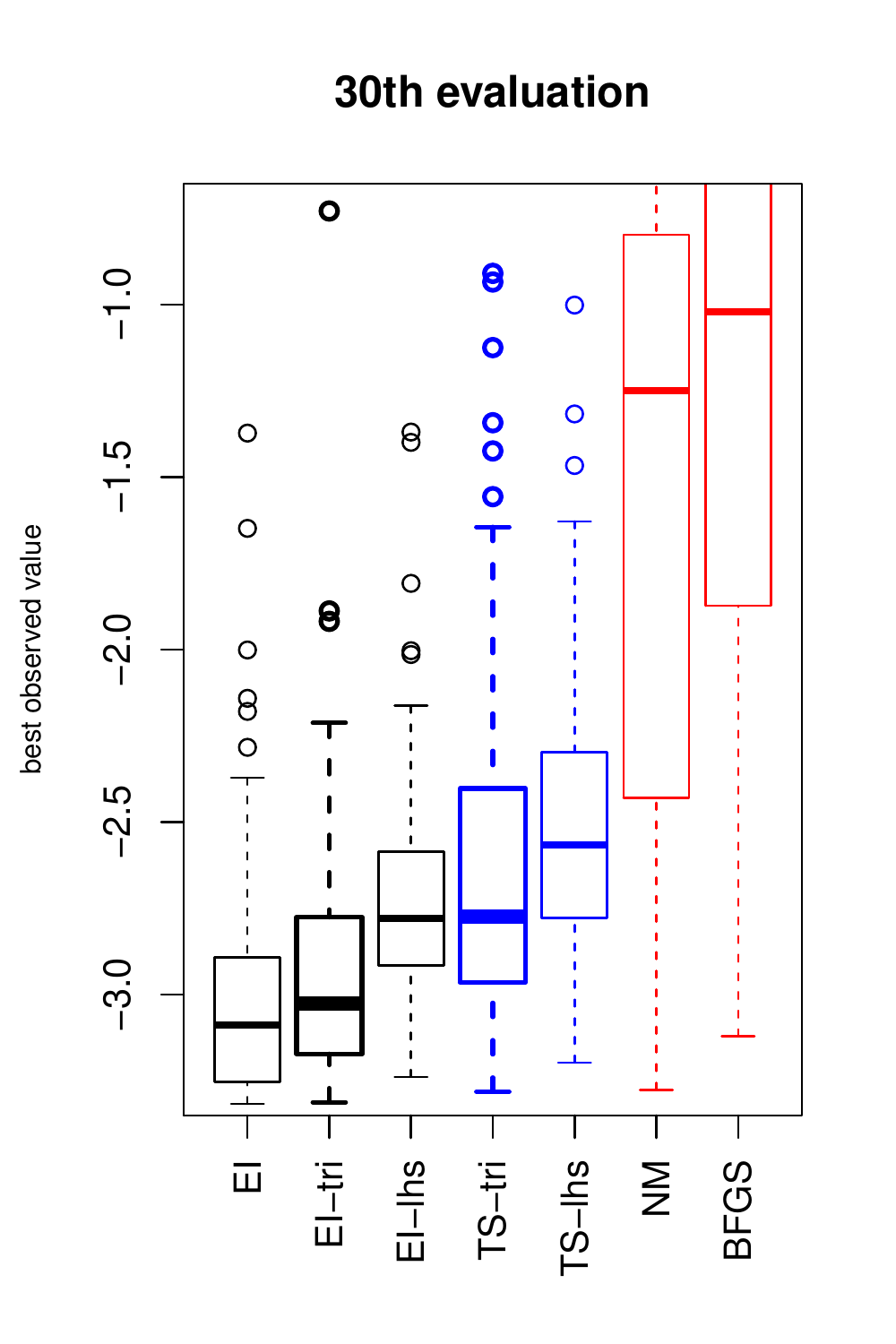} 
\includegraphics[scale=0.44,trim=50 10 20 10,clip=TRUE]{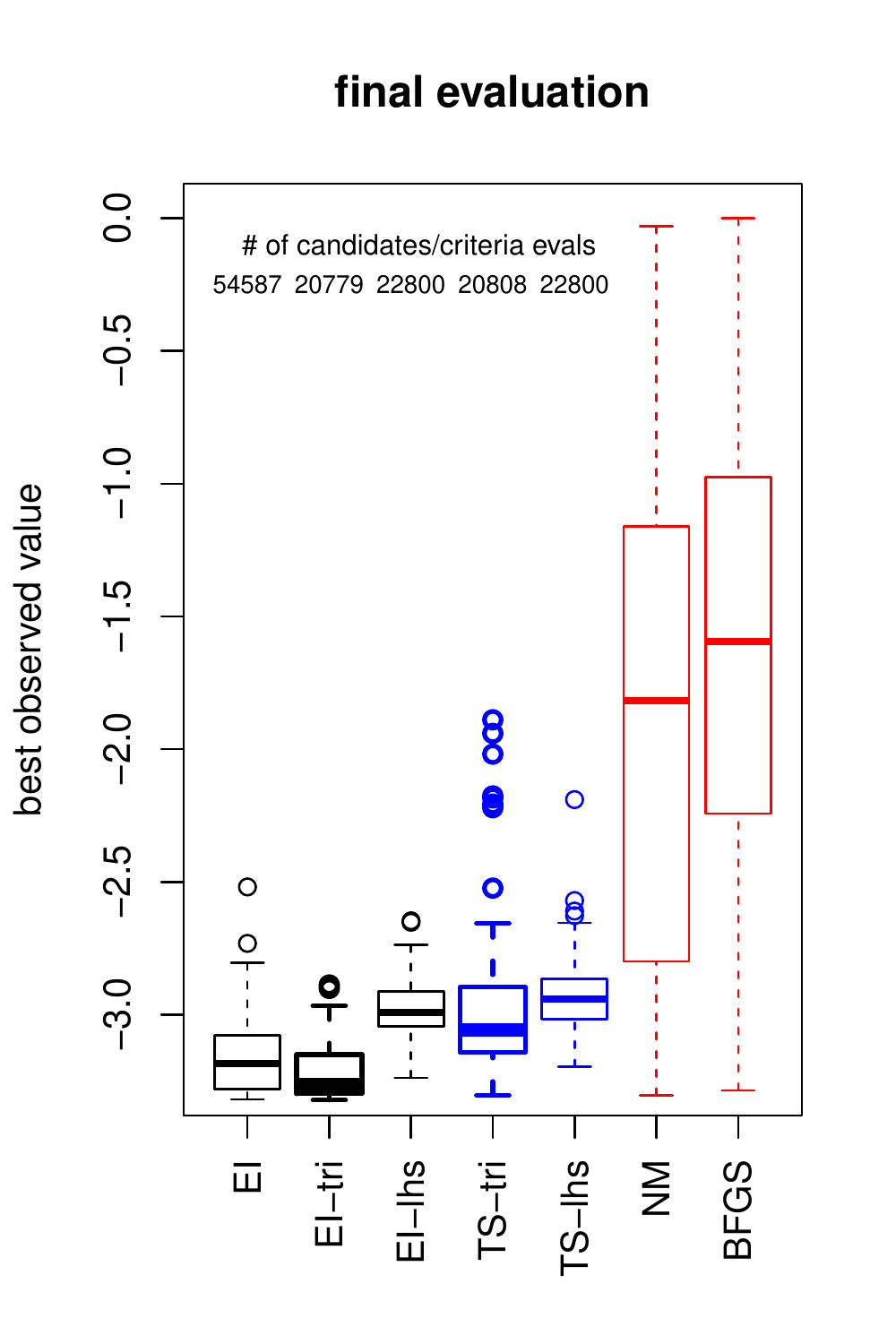} 
\vspace{-0.25cm}
\caption{Analog of Figure \ref{fig:gphart} for the Hartmann 6 function.
\label{fig:hartmann6}}
\end{figure*}

As a second example of conventional GP-based BO, consider the six-dimensional
Hartmann function. %\footnote{\url{http://www.sfu.ca/~ssurjano/hart6.html}}   
Again, see \citet{picheny2012benchmark} for more on this benchmark. Our
experimental setup is identical the Goldstein--Price experiment in Section \ref{sec:classic} except here we use the default $N_{\mathrm{sub}} = 100d = 600$. 
Figure \ref{fig:hartmann6} summarizes
results. Even with this high $N_{\mathrm{sub}}$, tricands still result in many
fewer acquisition criteria evaluations than numerically optimized EI (right panel),
which must search more aggressively for the local optimum in
such high dimension.

The results are broadly similar to our earlier 2d Goldstein--Price example
shown in Figure \ref{fig:gphart}. Tricands-based EI yields equivalent, or
better, BOV despite many fewer criteria evaluations.  This is true, but to a
lesser extent, with TS. It is notable that tricands came from behind in the
case of EI.  For the first twenty or so acquisitions, numerically optimized EI
bests its tricands analog.  Geometric bias towards exploration
may be less desirable in early acquisitions, but pays off by the end of the
search.

\subsection{Michaelwicz}
\label{app:mich}

As a second example of abruptly changing regimes, continuing from Section
\ref{sec:abrupt}, consider the Michaelwicz
function. %\footnote{\url{http://www.sfu.ca/~ssurjano/michal.html}} 
Like G\&L, the surface has large flat areas, but it also has a continuum of
ridges of local minima which intersect to create deeper valleys of local
minima, and ultimately one global minimum where the deepest of those ridges
intersect.  A nice feature of the Michaelwicz function is that it is defined
in arbitrary input dimension.  Here we use it in 4d, which makes for a very
difficult surface to model and optimize.  To cope, we take the search out to
$n_{\mathrm{end}} = 75$ and allow up to two hundred candidates per
acquisition.  Otherwise the setup is similar to earlier examples.

\begin{figure*}[ht!]
\centering
\includegraphics[scale=0.44,trim=5 10 25 10,clip=TRUE]{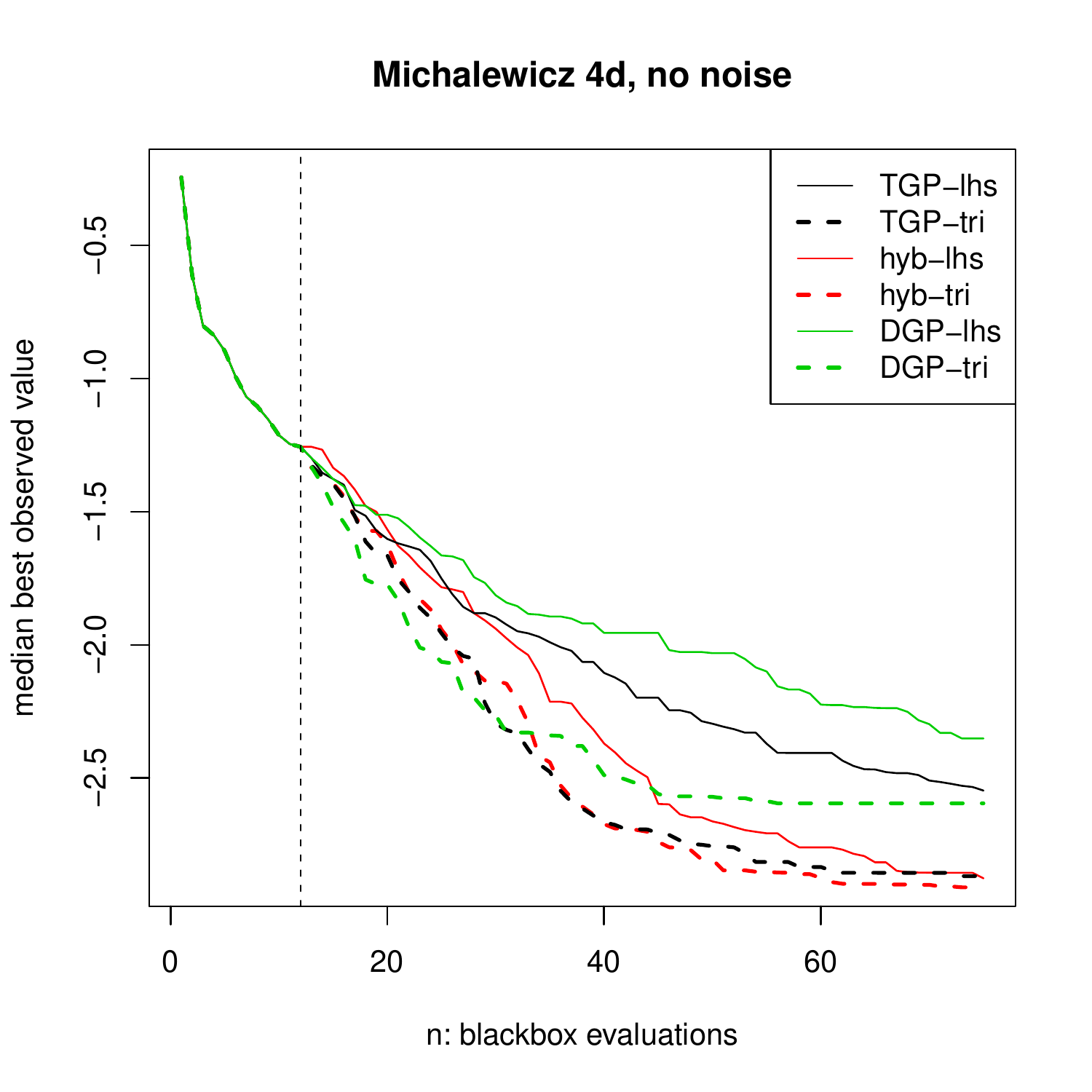}
\includegraphics[scale=0.44,trim=5 10 25 10,clip=TRUE]{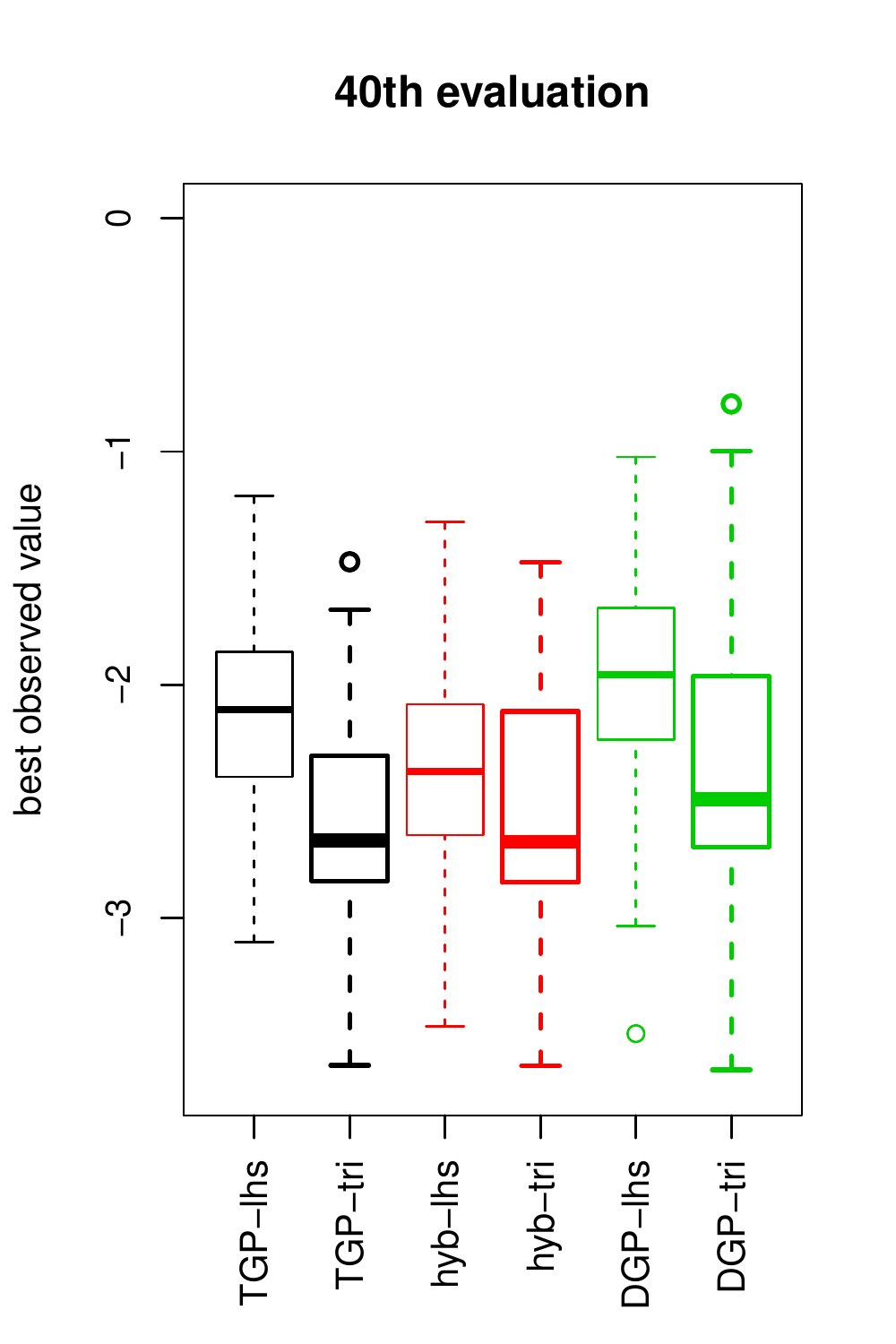} 
\includegraphics[scale=0.44,trim=50 10 20 10,clip=TRUE]{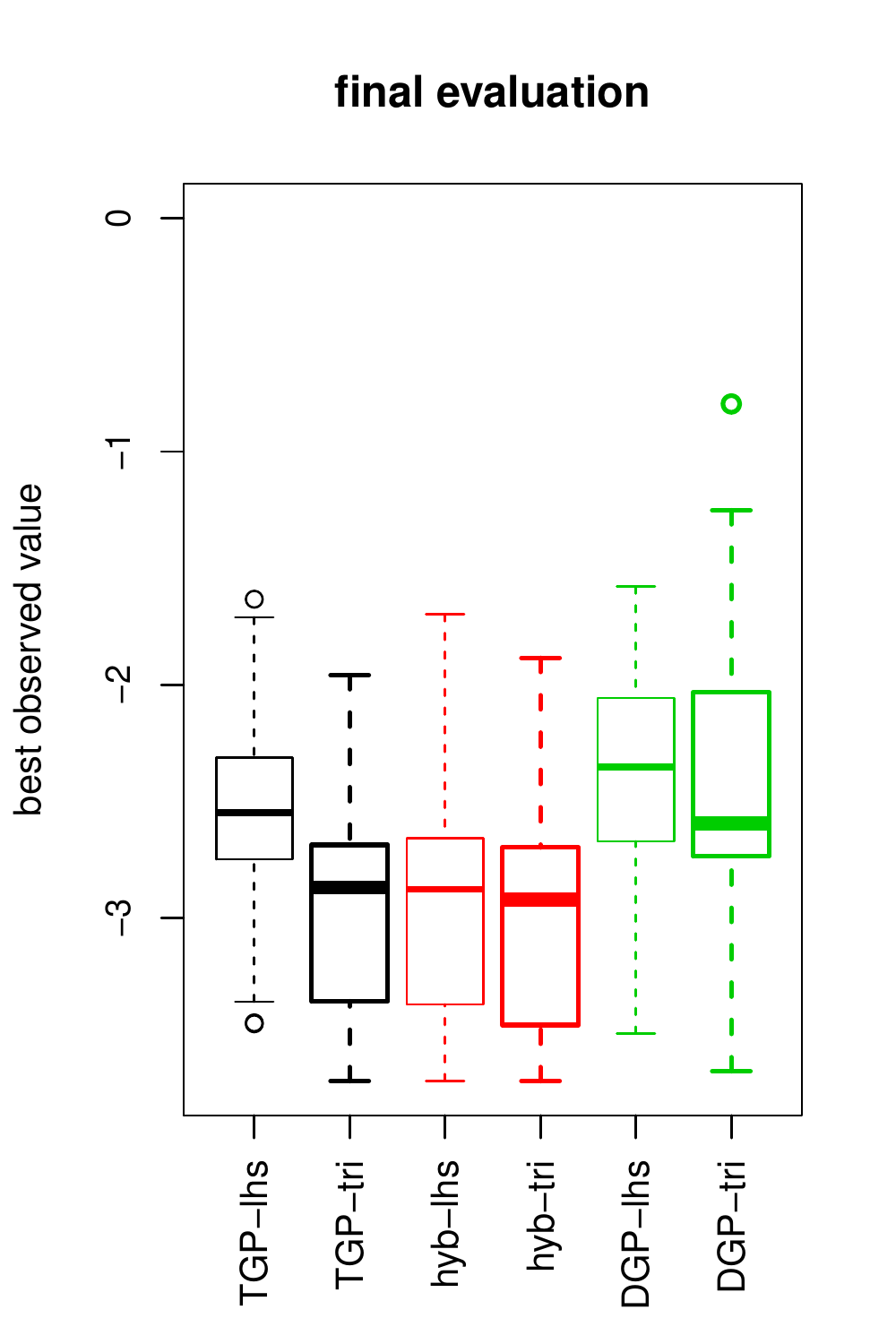} 
\vspace{-0.25cm}
\caption{Michaelwicz function in 4d with MCMC-based comparators.
\label{fig:mik} }
\end{figure*}

Figure \ref{fig:mik} shows the results, which are largely similar to G\&L,
except revealing of these additional challenges.  Tricands uniformly dominate
LHS via median BOV.  Although noise is high across all boxplots, reflecting
variability in solution quality over random re-initialization, the patterns
are clear in a pairwise analysis.  For example, the red boxplots in the right
panel look similar but, at the 75th evaluation, 75/100 tricands-based BOVs
were below their LHS counterpart.  As in the G\&L example, TGP bests DGP.
%Although tricands helps, progress seems to saturate after a while.  
Abrupt regime changes are at odds with the DGP's smooth warping of inputs.

\section{Additional discussion}
\label{app:discuss}

When reflecting on tricands' value, it is important to remember some stylized
facts about BO.  One is that the theory (under stringent regularity
conditions) guarantees convergence to the global minimum ``eventually,'' in
the sense that with enough samples you'll explore everywhere
\citep[e.g.,][]{bull2011convergence}. That's not much help in practice,
because exploring everywhere isn't practical. Another is that EI and TS are
greedy; their scope is the next acquisition.  Contemplating a remaining budget
and entertaining optimal decisions through that lens can be caricatured as a
herculean effort with marginal gains
\citep{gonzalez2016glasses,frazier2008knowledge,lam2016bayesian,gramacy2011optimization}.
These are not bad ideas, but they haven't moved the needle on the {\em modus
operandi} because they've not been incorporated into accessible libraries,
and are too involved for bespoke implementation in practice.

Both theory and aggressively scoped acquisition can lose sight of the real
goal. What's important in BO, and active learning in general, is creating a
virtuous cycle between data acquisition and learning.  Anyone who has tried
knows that there is a fine line between vicious and virtuous when it comes to
implementation details, despite the best of intentions and theoretical
``guarantees''. EI and TS are simple to implement because GP libraries are in
abundance, code evaluating criteria is a few lines long, and there are many
examples to cut-and-paste from.  Barriers to application are low, but it's
easy to get carried away to disappointment. This is what makes tricands
attractive.  It is motivated by simple principles: the solution is in-between
your current runs, so look there.  It is plug-n-play wherever candidates are
an option.  This means they can be applied in situations where numerical
differentiation is not available (e.g., with MC integrated surrogate).
Evaluation of the acquisition criteria can be massively parallelized with
candidates, say on a GPU, whereas local numerical solvers are inherently sequential.

That's not to say that tricands are a panacea.  Some of our boxplots indicated
that improvements over alternatives on best-observed-value (BOV) in 90/100 MC
repetitions may have come at the expense of the worst case performance of the
remaining 10\%. This may have simply been bad luck.  But if not, the
deterministic nature of tricands could be to blame: not enough opportunity to be
surprised.  As mentioned in the main body of the paper, one can always augment
tricands with random/space-filling candidates.  This could be especially
beneficial when $N$, the number of tricands for a given $n$, is lower than the
desired budget of candidates.  Entertaining tuning parameters for some of our
hard-coded settings, like the distance between fringe candidates and the
boundary (Section \ref{sec:fringe}), could help. That distance could even be
chosen at random.  We could likewise randomly choose a location for interior
candidates (Section \ref{sec:interior}), within each triangle, rather than
taking the barycenter.  When randomly downsampling (Section \ref{sec:subset}), we
could guarantee a certain proportion of fringe candidates like we did for ones
adjacent to $f_n^{\min}$.

Although the important subroutines of Delaunay triangulation and convex hulls
are off-loaded to libraries, they can (at times) be computationally demanding.
When $n$ is large and $d$ is modest, calculating $N$ locations in the
thousands (see Figure \ref{fig:size}) could be cumbersome.  Of course, the
whole goal of BO is to limit $n$.  In our experiments with $n \leq 75$, all of
our triangulation/hull calculations took fractions of a second. But with big
$n$ they can take minutes (Figure \ref{fig:tricands}, right panel), and that
could be prohibitive.  However, after each acquisition the number of new sites
only increases by one ($n\rightarrow n+1$), and thus affects only a small,
local part of the triangulation/hull. The {\tt Qhull} library does not support
this, but there are incremental algorithms for triangle/hull augmentation
which are very fast relative to starting from scratch.  For more details, see
\citet{su1997comparison}.

Such a strategy might be valuable in higher dimensional BO settings.
Continuous optimization theory tells us that gradient-based methods have local
convergence rates independent of dimension, which has made L-BFGS-B and
similar optimizers the tool of choice in solving for acquisitions. This,
together with the exponential growth of input space volume, might at first
blush suggest that tricands' performance is limited to low/modest dimension.
However, in practice, the highly nonconvex nature of the acquisition surface
significantly cheapens the theoretical results associated with gradient-based
optimization.   Indeed, recent work has achieved state-of-the-art performance
in high $d$ using candidate sets focused within a certain region of the input
space \citep{eriksson2019scalable,wang2020learning,daulton2021multiobjective},
though still built on traditional space-filling points such as LHS. It would
be interesting to see whether replacing these space-filling points with
tricands would be as beneficial in that setting as we have found it to be in
modest $d$. Furthermore, a popular approach in scaling BO to high dimension is
to reduce the input space by screening input variables or finding linear or
nonlinear embedding spaces, rendering the problem a low dimensional one (see
\citet{binois2021survey} for an overview). The aim of such an approach is in
part to make solving the acquisition problem easier, and there's no reason to
believe this wouldn't extend to tricands.

Our surrogates were GP-centric, extended to handle non-stationarity via treed
partitioning and smooth (deep GP) input-warping.  It would be interesting to
explore the value of tricands paired with more unconventional surrogates based on
trees, such as random forests \citep{breiman2001random} or tree-structure
Parzen estimators \citep{bergstra2011algorithms}, where inner-optimization via
gradient-based local search is a non-starter.  We presented results with EI
and TS-based acquisition criteria, and of course there are a litany of other
heuristics. Our early experiments additionally included the upper-confidence
bound \citep[UCB;][]{srinivas2009gaussian} and probability of improvement (PI)
criteria.  The former, for most settings of the tuning parameter, mirrored our
EI results whereas the latter was dominated by EI.  To reduce clutter, we
decided not to include them in our presentation here.

Our ATO example in Section \ref{sec:ato} involved a stochastic simulator with
input-dependent noise. Surrogate modeling and active learning for stochastic
simulation is still very much on the frontier of the computer experiments
landscape \citep{baker2020analyzing}. In such settings, the acquisition space
should be extended to include the possibility of obtaining a replicate run,
exactly duplicating one of the $n$ existing design elements
\citep{binois2018replication}. Replication can be advantageous in separating
signal from noise in generic active learning tasks, and specifically in the
context of BO \citep[][Section 4.2]{JSSv098i13}. Rather than entertaining a
hybrid search between a continuum of novel locations and a discrete set of
replicate sites, tricands could be leveraged to make the entire set of
candidates discrete, vastly simplifying the inner optimization search.

Batch acquisition, acquiring several new runs at once, is a common paradigm
in some settings.  Ideas with modified EI go back at least to
\citet{ginsbourger2007multi}  with several following thereafter
\citep[e.g.,][]{taddy2009bayesian,chevalier2013fast}. One, more recent
approach involves penalizing regions of the input space near earlier
acquisitions in the batch  \citep{gonzalez2016batch}.  This spirit could be
ported to a tricands setting: simply rule out any candidates which are in
triangles $T_j$ adjacent to those acquired earlier in the batch. 

Finally, although we have emphasized BO, other active learning criteria could
benefit from candidates well-spaced relative to the current design. Perhaps
the most common is for active learning targeting reduced integrated mean-squared prediction
error \citep[IMSPE,
e.g.,][]{leatherman2017computer,binois2018replication,zhang2020batch}. High
input variance locations, such as between design sites, are natural candidates
for reducing IMSPE. Another recently popular active learning topic in computer
experiments is contour/level set finding
\citep{ranjan2008sequential,bect2012sequential,chevalier2014fast,marques2018contour,azzimonti2020adaptive,cole2021entropy}.
Many of the criteria suggested in these works involve predictive entropy from
a GP surrogate, which is famously myopic; entropy (defined via a
classification above and below a level set) tends to be higher near training
data already near partition boundaries, leading to a clumping of acquisitions
unless explicit measures are taken to spread out candidates, or to otherwise
deter a numerical inner-optimizer.  Tricands could offer a geometric spread of
future acquisitions away from existing training data locations.

\end{document}